\documentclass[prd,aps,eqsecnum,floatfix,nofootinbib,preprint,tightenlines]{revtex4}

\usepackage{latexsym}
\usepackage{graphicx}
\usepackage[dvipsnames]{xcolor}

\def\bibi{\bibitem}

\let\inodot=\i

\def\a{\alpha}
\def\b{\beta}
\def\c{\chi}
\def\d{\delta}
\def\f{\phi}                    
\def\g{\gamma}

\def\i{\iota}

\def\m{\mu}
\def\n{\nu}
\def\o{\omega}
\def\p{\pi}                     
\def\r{\rho}                    
\def\s{\sigma}                  
\def\t{\tau}

\def\D{\Delta}

\def\P{\Pi}

\def\cbo{{\,\raise-.15ex\Sc [\,}}                       
\def\ltap{\raisebox{-.4ex}{\rlap{$\sim$}} \raisebox{.4ex}{$<$}}   
\def\gtap{\raisebox{-.4ex}{\rlap{$\sim$}} \raisebox{.4ex}{$>$}}   

\def\ddt#1{{\buildrel {\hbox{\LARGE .\kern-2pt.}} \over {#1}}}

\def\ie{\mbox{\it i.e.}}
\def\eg{\mbox{\it e.g.}}
\def\etc{\mbox{\it etc.}}

\def\floatcaption#1#2{ \caption{ #2 \ [#1] \label{#1}} }
\def\floatcaption#1#2{ \caption{#2 \label{#1}} }

\long\def\symbolfootnote[#1]#2{\begingroup%
\def\thefootnote{\fnsymbol{footnote}}\footnote[#1]{#2}\endgroup}

\long \def \blockcomment #1\endcomment{}

\def\seef{{\it cf.}}

\def\bu{{\overline{u}}}
\def\bd{{\overline{d}}}
\def\bs{{\overline{s}}}

\def\ansatz{{\it ansatz}}

\begin{document}

\thispagestyle{empty}

\hfill KEK-TH-2049, LTH 1156
\begin{center}
\vspace*{5mm}
\begin{boldmath}
{\large\bf The strong coupling from $e^+e^-\to$~hadrons below charm}
\end{boldmath}
\\[10mm]
Diogo Boito,$^a$
Maarten Golterman,$^{b,c}$ Alexander Keshavarzi,$^d$
Kim Maltman,$^{e,f}$ Daisuke Nomura,$^g$
Santiago Peris,$^b$ Thomas Teubner$^d$
\\[8mm]
{\small\it
$^a$Instituto de F{\'\inodot}sica de S{\~a}o Carlos, Universidade de S{\~a}o Paulo\\
CP 369, 13570-970, S{\~a}o Carlos, SP, Brazil
\\[5mm]
$^b$Department of Physics and IFAE-BIST, Universitat Aut\`onoma de Barcelona\\
E-08193 Bellaterra, Barcelona, Spain
\\[5mm]
$^c$Department of Physics and Astronomy,
San Francisco State University\\ San Francisco, CA 94132, USA
\\[5mm]
$^d$Department of Mathematical Sciences, University of Liverpool, Liverpool L69 3BX, U.K.
\\[5mm]
$^e$Department of Mathematics and Statistics,
York University\\  Toronto, ON Canada M3J~1P3
\\[5mm]
$^f$CSSM, University of Adelaide, Adelaide, SA~5005 Australia
\\[5mm]
$^g$KEK Theory Center, Tsukuba, Ibaraki 305-0801, Japan}
\\[10mm]
\end{center}

\begin{quotation}
We use a new compilation of the hadronic $R$-ratio from available data for
the process $e^+e^-\to\mbox{hadrons}$ to determine the strong coupling, $\a_s$.
We make use of all data for the $R$-ratio from threshold to a center-of-mass energy
of 2~GeV by employing
finite-energy sum rules.   Data above 2~GeV, for which at present far
fewer high-precision experimental data are available,
do not provide much additional constraint but are 
fully consistent with the values for $\a_s$ we obtain.   
Quoting our results at the $\t$ mass  to facilitate comparison to the results obtained from 
analogous analyses of hadronic $\t$-decay data, 
we find $\a_s(m_\t^2)=0.298\pm 0.016\pm 0.006$
in fixed-order perturbation theory, and $\a_s(m_\t^2)=0.304\pm 0.018\pm 0.006$
in contour-improved perturbation theory, where the first error is statistical, and the
second error reflects our estimate of various systematic effects.    
These values are in good agreement
with a recent determination from the OPAL and ALEPH data for
hadronic $\t$ decays.
\end{quotation}

\newpage
\section{\label{introduction} Introduction}
There are many hadronic quantities from which the strong coupling, $\a_s(s)$, can be extracted,
at many different energy scales $E=\sqrt{s}$, as long as $s$ is large enough that QCD perturbation
theory can be expected to apply.   The range of scales employed in such determinations ranges from
above the $Z$ mass, where non-perturbative effects are negligble, down to the $\t$ mass,
where these effects, although subdominant, must be taken into account carefully in an accurate extraction of $\a_s$.  
Not all of these determinations lead to values for $\a_s$ (quoted, for instance, in the 5-flavor, $\overline{\rm MS}$ scheme at the $Z$ mass) that are competitive when comparing the errors.\footnote{For a recent review, see Ref.~\cite{salam}.}
Nevertheless, determinations over a wide range of scales are interesting, because they 
directly test the running of the coupling predicted by QCD.   As such, it is interesting to consider
determinations of $\a_s$ at scales as low as the $\t$ mass.   

Some years ago, a calculation of the five-loop contribution to the Adler function \cite{PT} revived
interest in the determination of $\a_s$ from non-strange hadronic $\t$ decays; for recent work
see Refs.~\cite{BJ,MY08,CF,alphas1,alphas2,Cetal,ALEPH13,alphas14,Pich,BGMP16}.   The results 
of these efforts have been controversial,\footnote{See, in particular, Refs.~\cite{Pich,BGMP16} for a
clear account of the controversy.} because it is difficult to disentangle non-perturbative 
contributions to the spectral functions extracted from hadronic $\t$ decays, 
and, in fact, it is not obvious that this can be done in a completely satisfactory way.
Moreover, it is difficult to make progress in the context of hadronic $\t$ decays, because the $\t$
mass puts a limit on the scales that can be probed within this approach.

It would thus be interesting to apply and test the same techniques in a similar setting where 
no such limit exists.   This leads us to consider, instead of $\t$ decays, the $R$-ratio
$R(s)$, measured in the process $e^+e^-\to\mbox{hadrons}(\g)$, which is directly 
proportional to the electromagnetic (EM) QCD vector spectral function.\footnote{The symbol
$(\g)$ indicates that the hadronic final state is inclusive of final-state radiation.}   The same
technology used in extracting $\a_s$ from the non-strange, $I=1$, vector and axial
spectral functions measured in hadronic $\t$ decays can also be used to extract $\a_s$
from the EM spectral function.   The technology used in $\t$ decays, which we apply here to
$R(s)$ instead, is that of finite-energy sum rules (FESRs) \cite{shankar,Braaten88,BNP}.

The idea of comparing the predictions from QCD perturbation theory with $R(s)$ at large enough
$s$ is an old and obvious one.   However, the extraction of $\a_s(s)$ from $R(s)$ at a single value
of $s$ leads to a very large uncertainty, which makes the resulting $\a_s$ compatible with other extractions, but uninteresting as a source of precise information about the coupling.\footnote{See,
for instance, Refs.~\cite{BES,BES3,KEDR}, in particular, Table 3 in Ref.~\cite{BES3}.}
The use of FESRs, instead, allows us to make use of all data for
$R(s)$ from threshold to some $s=s_0$, to extract $\a_s$ with a much 
higher precision than can be obtained from a ``local'' determination at the scale $s=s_0$.   
The reason an FESR determination is expected to be more precise is that, rather than relying only on a single local $R(s)$ result, FESRs employ weighted integrals over the experimental spectral distribution 
for $s$ running from threshold to some upper limit $s_0$. Since the experimental data are 
more precise at 
lower $s$, the weighted spectral integrals for $s_0$ in the region where $R(s)$ starts to behave
perturbatively are typically much more precise than are the values of $R(s)$  in the same region. The associated FESR determinations of $\a_s$ are thus also expected to be much more precise than those 
obtained by matching the perturbative expression for $R(s)$ to the spectral data directly.   As we will see, a new compilation of $R(s)$ combining all available  experimental electroproduction cross-section results
\cite{KNT18} makes it possible to determine $\a_s$ at scales $s_0$ for  $m_\t^2\,\ltap\, s_0\leq 4$~GeV$^2$ with an error small enough to make the comparison with other determinations of $\a_s$
interesting.   Moreover, we expect that future, more precise data for $R(s)$ will allow us
to improve this determination of $\a_s$, because at present
the errors turn out to be dominated by those coming from the experimental errors on $R(s)$.

As this paper will show, it is the data for the FESR integrals over $R(s)$ up to $s_0$ for 
values between $s_0\approx m_\t^2$ and
$s_0=4$~GeV$^2$ that will contribute most to the accuracy with which we can determine
$\a_s$.   Of course, data for $R(s)$ beyond 4~GeV$^2$ exist, but their
accuracy is not yet sufficient to have a significant impact on the error in the determination
of $\a_s$.   Although the $\t$ mass plays no physical role in the current analysis, we will nonetheless
quote our $n_f=3$ flavor results for $\a_s$ at the scale $\m =m_\t$ in order
to facilitate direct comparison to the results of the analogous $\t$-based analyses.

The controversies that have plagued the determination of $\a_s$ from $\t$ decays are 
primarily related to the need to model violations of quark-hadron duality associated with
the clearly visible effects of hadronic resonances in the vector and axial spectral functions for
$s\le m_\t^2$.   At energies beyond the $\t$ mass, duality violations are expected to decrease
exponentially, making this a  major motivation for considering the determination of
$\a_s$ from $e^+e^-\to\mbox{hadrons}(\g)$.   Indeed,
while resonance effects are still present in the region $m_\t^2\le s\le
4$~GeV$^2$, it turns out that our central value for $\a_s$ from $R(s)$ is much less sensitive to 
the treatment of residual duality violations than was the case for $\t$-based analyses, 
with the modeling of these effects only needed
as part of the analysis of systematic errors.    It turns out that, given the current experimental
errors on $R(s)$, our estimate for the systematic error due to duality violations is rather small.

This paper is organized as follows.   In Sec.~\ref{theory}, we provide a brief review of the necessary
theory of FESRs.   Contributions from perturbation theory (in the $\overline{\rm MS}$ scheme) and the operator product expansion (OPE) are discussed in Sec.~\ref{OPE} and the inclusion of electromagnetic corrections in the OPE (necessitated by the fact that the hadronic final states include photons) in Sec.~\ref{EM}. The contributions from duality violations are considered in some detail 
in Sec.~\ref{DV}.   We describe and discuss the data in Sec.~\ref{data}, before turning to our analysis
in Sec.~\ref{analysis}.   Section~\ref{fits} contains our main fits to the data, Sec.~\ref{tests}
discusses systematic errors, and Sec.~\ref{results} contains our results, including a conversion to
the five-flavor $Z$-mass scale. In Sec.~\ref{tau} we compare these results to those obtained from an analogous $\t$-based analysis.   Section~\ref{conclusion} contains our conclusions.

\section{\label{theory} Theory}
In this section, we review the FESR methodology, as applied to the case of the two-point function
of the three-flavor EM current, 
\begin{equation}
\label{EMcurrent}
J^{\rm EM}_\m=\frac{2}{3}\,\bu\g_\m u-\frac{1}{3}\,\bd\g_\m d-\frac{1}{3}\,\bs\g_\m s
=J^{3}_\m+{\frac{1}{\sqrt{3}}}J^8_\m\ ,
\end{equation}
where the superscripts $3$ and $8$ label the neutral $I=1$ and $I=0$ members of the $SU(3)$ 
octet of three-flavor vector currents, respectively.  The EM vacuum polarization 
$\P(q^2)$ is defined through\footnote{Note that,
with this definition, in the isospin limit, the $I=1$ part
of $\P(q^2)$ has a normalization one-half that of the corresponding
isovector flavor $ud$ polarization encountered in the analysis of
hadronic $\tau$ decays.}
\begin{eqnarray}
\label{EMpol}
\P^{\rm EM}_{\m\n}(q)&=&i\int d^4x\,e^{iqx}\langle 0|T\left\{J^{\rm EM}_\m(x)J^{\rm EM}_\n(0)\right\}|0\rangle\ ,\\
&\equiv &\left(q_\m q_\n-q^2 g_{\m\n}\right)\P(q^2)\ ,\nonumber
\end{eqnarray}
and the corresponding spectral function is obtained, as usual, from the imaginary part of $\P(q^2)$ as\footnote{We will drop the superscript EM on $\P(q^2)$.}
\begin{equation}
\label{spec}
\r(s)=\frac{1}{\p}\,\mbox{Im}\,\P(s)=\frac{1}{12\p^2}\,R(s)\ .
\end{equation}
The second equality in Eq.~(\ref{spec}) follows from the fact that the imaginary part of $\P(q^2)$
is directly related to the cross section for $e^+e^-\to\mbox{hadrons}$, through the optical theorem.
Here $R(s)$ is defined by
\begin{equation}
\label{R}
R(s)\equiv\frac{3s}{4\p\a^2}\,\s_{e^+e^-\to{\rm hadrons}(\g)}(s)
=\frac{\s_{e^+e^-\to{\rm hadrons}(\g)}(s)}{\s_{e^+e^-\to\m^+\m^-}(s)}\ ,
\end{equation}
where $\a$ is the fine-structure constant, and the second equation holds for values of
$s$ for which we can neglect
the muon mass.   The $\g$ in parentheses indicates that hadronic states with final-state radiation
are included in addition to purely hadronic states.

In Sec.~\ref{FESR} we review the FESRs which relate $R(s)$, which is available
from experimental data for $e^+e^-\to\mbox{hadrons}$, to a theoretical representation of 
$\P(q^2)$ at large $q^2$.   In Secs.~\ref{OPE} and~\ref{EM} we review the theoretical representation for
large $q^2$ away from the Minkowski axis $q^2=s$, based on the OPE.   As is well known,
the OPE does not capture the non-analytic behavior of $\P(q^2)$ on the positive
real $q^2$ axis that corresponds to the presence of hadronic resonances in $\r(s)$.   In Sec.~\ref{DV} we 
discuss our method for modeling these ``duality-violating'' 
effects, and the use of this approach in estimating the systematic
uncertainty associated with neglecting duality-violating effects in the
determination of $\a_s$ from FESR analyses of $\P(q^2)$.

\subsection{\label{FESR} Finite-energy sum rules}
\begin{figure}[t]
\vspace*{4ex}
\begin{center}
\includegraphics*[width=6cm]{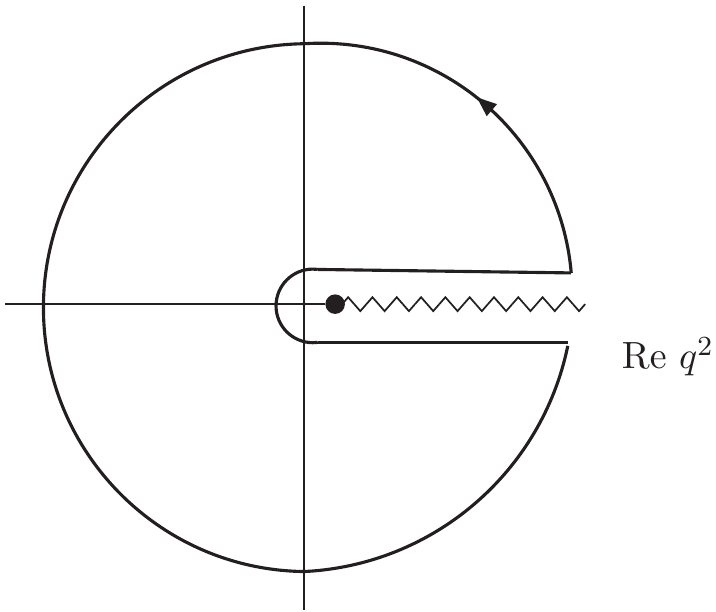}
\end{center}
\begin{quotation}
\floatcaption{cauchy-fig}%
{{\it Analytic structure of $\P(q^2)$ in the complex $z=q^2$ plane.
There is a cut
on the positive real axis starting at $s=q^2=m_\p^2$ (see text).
The solid curve shows the contour used in Eq.~(\ref{cauchy}).}}
\end{quotation}
\vspace*{-4ex}
\end{figure}
Extending $z=q^2$ to the complex plane, the function $\P(z)$ is analytic everywhere except 
on the positive real $z$-axis.   Therefore, the integral of $\P(z)$ times any analytic function
of $z$, along the contour shown in
Fig.~\ref{cauchy-fig}, vanishes.   From this, employing Eq.~(\ref{spec}), one has, for any polynomial 
weight $w(y)$, the FESR relation
\begin{equation}
\label{cauchy}
I^{(w)}(s_0)\equiv\frac{1}{12\p^2 s_0}\int_0^{s_0}ds\,w\left(\frac{s}{s_0}\right)\,R(s)
=-\frac{1}{2\p i\, s_0}\oint_{|z|=s_0}
dz\,w\left(\frac{z}{s_0}\right)\,\P(z)\ .
\end{equation}
We will use experimental data for $R(s)$ to evaluate the integrals on 
the left-hand side of Eq.~(\ref{cauchy}).   As already indicated in Eq.~(\ref{R}), these data also include
EM corrections, and the threshold value of $s$ is thus equal to $m_\p^2$, corresponding to the 
opening of the channel $e^+e^-\to\p^0\g$.     

In this paper, we will consider the weights
\begin{eqnarray}
\label{weights}
w_0(y)&=&1\ ,\\
w_2(y)&=&1-y^2\ ,\nonumber\\
w_3(y)&=&(1-y)^2(1+2y)\ ,\nonumber\\
w_4(y)&=&(1-y^2)^2\ ,\nonumber
\end{eqnarray}
where the subscript indicates the degree of the polynomial.
  The weight $w_2(y)$ has a single zero at $z=s_0$ (a single ``pinch''), suppressing contributions from the region near the timelike point $z=s_0$ on the contour.   The weights $w_3(y)$ and $w_4(y)$ are doubly pinched,
with a double zero at $z=s_0$.   All weights are chosen such that no linear term in $y$ appears;
the reason for this is discussed in the next section.
The weights~(\ref{weights}) form a linearly independent basis for polynomials up to degree four
without a linear term. 

\subsection{\label{OPE} Perturbation theory and the OPE}
We begin with splitting $\P(z)$ into two parts:
\begin{equation}
\label{splitPi}
\P(z)=\P_{\rm OPE}(z)+\left[\P(z)-\P_{\rm OPE}(z)\right]\equiv \P_{\rm OPE}(z)+\D(z)\ ,
\end{equation}
where $\P_{\rm OPE}(z)$ is the OPE approximation to $\P(z)$,
\begin{equation}
\label{OPEdef}
\P_{\rm OPE}(z)=\sum_{k=0}^\infty\frac{C_{2k}(z)}{(-z)^k}\ .
\end{equation}
We will return to $\D(z)$ in Sec.~\ref{DV}.
Each of the coefficients $C_{2k}(z)$, for $k>1$, is a sum over contributions from different condensates
of dimension $D=2k$.   The
$D=0$ term corresponds to the purely perturbative contribution obtained
in massless perturbation theory; the $D=2$ term to the perturbative 
contributions proportional to the squares of the light quark masses.
Each contribution depends logarithmically on $z$, and this dependence
can be calculated in perturbation theory.  In practice, it is convenient to consider, instead
of $\P(z)$, the Adler function $D(z) \equiv -zd\P(z)/dz$,
which is finite and independent of the renormalization scale $\m$. The $D=0$ contribution, $D_0(z)$, to $D(z)$ takes the form
\begin{equation}
\label{pertth}
D_0(z)\equiv -z\frac{dC_0(z)}{dz}=\frac{1}{6\p^2}\sum_{n=0}^\infty\left(\frac{\a_s(\m^2)}{\p}\right)^n\sum_{k=1}^{n+1}
kc_{nk}\left(\log\frac{-z}{\m^2}\right)^{k-1}\ ,
\end{equation}
where the coefficients $c_{nk}$ are known to five-loop order, \ie, order $\a_s^4$~\cite{PT}.
It is straightforward to rewrite the 
$D=0$ contributions to the right-hand side of Eq.~(\ref{cauchy}) in terms of $D_0(z)$ via partial
integration.  
The independence of $D(z)$ on $\m$ implies that only the
coefficients $c_{n1}$ are independent; the $c_{nk}$ with $k>1$
can be expressed in terms of the $c_{n1}$ through use of the renormalization group,
resulting in expressions also 
involving the coefficients of the $\b$ function.\footnote{See for instance Ref.~\cite{MJ}.}   In the
$\overline{\rm MS}$ scheme, $c_{01}=c_{11}=1$, $c_{21}=1.63982$, $c_{31}=6.37101$
and $c_{41}=49.07570$, for three flavors \cite{PT}.\footnote{In this paper, we will restrict ourselves to the $\overline{\rm MS}$
scheme, even though it may be interesting to investigate other ``physical'' schemes as well
\cite{BJM}.}   While $c_{51}$ is not currently known, we will use the estimate
$c_{51}=283$ provided
in Ref.~\cite{BJ}, to which we assign an uncertainty $\pm283$.    For the running of $\a_s$ we use the
four-loop $\overline{\rm MS}$ $\b$-function, but we have checked that using 5-loop running instead
\cite{5loop} leads to differences of order $10^{-4}$ or less in our results for $\a_s$ at the $\t$ mass.

Beyond the uncertainty in $c_{51}$, it is common practice to consider different guesses
about higher orders in perturbation theory, in order to obtain insight into the effect of 
neglecting terms beyond
those explicitly included in evaluating the $D=0$ contribution to the right-hand side of 
Eq.~(\ref{cauchy}).   Two commonly used prescriptions are fixed-order
perturbation theory (FOPT), in which  $\m$ is chosen to be a fixed scale, here $\m^2=s_0$, and
contour-improved perturbation theory (CIPT \cite{CIPT}), in which  the scale $\m^2$ is set 
equal to $-z$, thus resumming to all orders the running of the coupling point-by-point along the contour,
using the 4-loop beta function (so only terms with $k=1$ survive in Eq.~(\ref{pertth})).      
The two procedures lead
to different values of $\a_s$. This difference is a source of systematic uncertainty in this type of analysis.

We next turn to the quadratic, mass-dependent perturbative contributions
encoded in the $D=2$ term, $C_2(z)$, of Eq.~(\ref{OPE}).   
With terms proportional to the squares of the light quark masses $m_{u,d}$ safely negligible,
 $C_2(z)$ is proportional to $m^2_s$, the square of the strange quark mass, and takes the form
\begin{equation}
\label{D=2}
C_2(z)=\frac{m_s^2(\m^2)}{6\p^2}\sum_{n=0}^\infty\left(\frac{\a_s(\m^2)}{\p}\right)^n\sum_{k=0}^n
f_{nk}\left(\log\frac{-z}{\m^2}\right)^k\ .
\end{equation}
By choosing $\m^2=-z$, one recovers the 
result derived in Refs.~\cite{ChK93}, with $f_{00}=1$, $f_{10}=8/3$ and $f_{20}=23.26628$,
truncating the series at three-loop order.   Here we will use the fixed-order expression with
$\m^2=s_0$ in Eq.~(\ref{D=2}).  The coefficients $f_{nk}$ with $k>0$ can again be expressed
in terms of the $f_{n0}$ 
by using the renormalization group; they involve the coefficients of the $\b$ function and the
mass anomalous dimension $\g$.   With the $D=2$ 
contribution representing a small correction to the $D=0$ 
term,\footnote{Its presence shifts the value of $\a_s$ by about
1-2\%.} the impact on the values of $\a_s$ obtained in our analysis of a shift 
from the fixed-order to contour-improved scheme for treating the
$D=2$ contribution is safely negligible.\footnote{The treatment of the rather similar $D=2$ OPE 
series for the flavor $ud-us$ $V+A$ polarization, which is obtained from that in Eq.~(\ref{D=2}) after 
rescaling by $9$ and setting $f_{00}=1$, $f_{10}=7/3$, $f_{20}=19.93$ and $f_{30}=208.75$, 
has been studied by comparing lattice and OPE results in Ref.~\cite{hlmz17}. The results of that study 
favor the use of 3-loop truncation and the FOPT scheme.
It is thus reasonable to expect these choices to be optimal here as well.}
We will run the strange quark mass to the scale $s_0$ from
$\m=m_\t$, employing the $\overline{\rm MS}$ value $m_s(m^2_\t)=97$~MeV
as input.\footnote{This corresponds to the $2+1$ flavor, $\overline{\rm MS}$ value $m_s(\m=2~\mbox{GeV})=92$~MeV,
taken from Ref.~\cite{FLAG3}.}

The $D=4$ term, $C_4(z)$, does not contribute to the sum rules~(\ref{cauchy}) if we ignore 
its logarithmic dependence on $z$, because none of the weights in Eq.~(\ref{weights})
contains a term linear in $y$.   The $z$ dependence 
for these weights enters the right-hand side of Eq.~(\ref{cauchy}) only at order $\a_s^2$.   These effects were found to be safely 
negligible in the analogous sum-rule analysis of hadronic $\t$ decay
data reported in Ref.~\cite{alphas1}.  Since in this
paper we will work at values of $s_0$ larger 
than those employed in the $\t$-based analysis, it is safe to neglect these 
effects here as well.   This means the $D=4$ term
plays no role in our analysis.   Our avoidance of sum rules involving the $D=4$ term
is motivated by the results of Ref.~\cite{BBJ12}, in which a renormalon-model-based study
indicated that perturbation theory for sum rules with such weights is particularly unstable.\footnote{Earlier considerations along the same lines can be found in Refs.~\cite{BJ,alphas1,MJ}.}

We will also ignore the logarithmic $z$ dependence of the higher-order coefficients $C_D$, with $D\ge 6$,
for the simple reason that no complete information on this dependence is available.   We note that,
of course, the $z$ dependence is again suppressed by a power of $\a_s$.
This means that the FESR with weight $w_2$ will involve $C_6$, the FESR with weight $w_3$
will involve $C_6$ and $C_8$, and the FESR with weight $w_4$ will involve $C_6$ and $C_{10}$.
The presence of $C_6$ in different sum rules provides an additional
consistency check on our fits.   As the OPE itself diverges as an expansion in
$1/z$, it is safer to include sum rules with low-degree weights such as $w_0$ and 
$w_2$ in the analysis.

\subsection{\label{EM} EM corrections}
Since the experimental data for $R(s)$ include EM corrections, we also have to incorporate such
corrections on the right-hand side of the sum rules~(\ref{cauchy}).   It turns out that the only
numerically significant correction is the leading-order correction to the $D=0$ term \cite{sug}
and, in our analysis, we thus correct the $n=0$ term in Eq.~(\ref{pertth}) by the replacement
\begin{equation}
\label{EMcorr}
\frac{1}{6\p^2}\,c_{01}\to \frac{1}{6\p^2}\,c_{01}\left(1+\frac{\a}{4\p}\right)\ ,
\end{equation}
where $\a$ is the fine-structure constant.
The numerical effect of this replacement is to shift the value for $\a_s(m_\t^2)$ obtained in our
analysis by about $-0.001$.   EM corrections subleading to the correction shown in Eq.~(\ref{EMcorr}) turn out to be completely irrelevant, numerically.

\subsection{\label{DV} Duality violations}
We next turn to the contribution of $\D(z)$, defined in Eq.~(\ref{splitPi}), to the 
sum rules~(\ref{cauchy}).   As shown in Refs.~\cite{CGP05,CGPmodel}, under the condition that the integral over $w(z/s_0)\D(z)$ around the circle with radius $s_0$ goes to zero for $s_0\to\infty$, this integral can be rewritten such that the sum rule takes the form
\begin{eqnarray}
\label{sumrule}
I^{(w)}(s_0)&=&-\frac{1}{2\p i\, s_0}\oint_{|z|=s_0}
dz\,w\left(\frac{z}{s_0}\right)\,\P_{\rm OPE}(z)-\frac{1}{s_0}\int_{s_0}^\infty ds\,w\left(\frac{s}{s_0}\right)\r_{\rm DV}(s)\ ,\nonumber\\
\r_{\rm DV}(s)&\equiv&\frac{1}{\p}\,\mbox{Im}\,\D(s)\ .
\end{eqnarray}
In this form, the origin of the extra term in the FESR becomes clear:  the duality-violating
part of the spectral function, $\r_{\rm DV}(s)$,
represents the part of the spectral function which is not captured by the OPE.   In 
physical terms, this results from the deviations from the monotonic OPE behavior resulting from the presence of resonances in the spectrum, for large $s$.  

Building on earlier work \cite{russians}, a framework for the understanding of duality violations
in terms of a generalized Borel-Laplace transform of $\P(q^2)$ and hyperasymptotics 
was developed in Ref.~\cite{BCGMP}.
Employing the $1/N_c$ expansion, working in the chiral limit, and assuming that for 
high energies the spectrum becomes Regge-like in the $N_c\to\infty$ limit, it was shown that, for
a given QCD channel, $\r_{\rm DV}(s)$ can be parametrized as
\begin{equation}
\label{ansatz}
\r_{\rm DV}(s)=e^{-\d-\g s}\sin(\a+\b s)\ ,
\end{equation}
for large $s$, up to slowly varying logarithmic corrections in the 
argument of the sine factor, and with $\g\sim 1/N_c$ small but non-zero.\footnote{This form was first used in Ref.~\cite{CGP05}, and subsequently further studied and employed in Refs.~\cite{alphas1,alphas2,alphas14,CGPmodel,CGP}.}   The parameter $\b$ is directly related to
the Regge slope, and the parameter $\g$ to the (asymptotic) ratio of the width and the mass
of the resonances in a given channel.  This form was sufficient for use in the case of
hadronic $\t$ decays, where we considered only the non-strange $I=1$ channel.\footnote{In
the case of $\t$ decays we took
the parameters in Eq.~(\ref{ansatz}) different in the vector and axial channels,
reflecting the differences in the resonance locations and widths in the two channels.}

Here, the situation is more complicated.   First, the EM current consists of two parts,
the $I=1$ and $I=0$ parts $J_\m^3$ and $J_\m^8$ of Eq.~(\ref{EMcurrent}), respectively.
Furthermore, it is not clear whether one can neglect the strange quark mass in the context
of duality violations, and use the chiral limit result 
Eq.~(\ref{ansatz}) for the strange quark component of the EM current.   For $m_s=0$, $SU(3)$
flavor symmetry implies that the duality violating parameters $\delta$, $\gamma$, $\beta$
and $\alpha$ in Eq.~(\ref{ansatz}) must be the same for the $I=1$ and 
$I=0$ channels.
However, the methods of Ref.~\cite{BCGMP} do not allow for a straightforward generalization to the
case of a non-zero quark mass, and this leaves us with the question as to how to parametrize
the $I=0$ part of $\r_{\rm DV}(s)$.

We will proceed as follows.   First, in considering duality-violation corrections, we will ignore 
disconnected contributions, which include
strange-light mixing, as this is doubly $SU(3)$-flavor and $1/N_c$ suppressed in the 
EM polarization.\footnote{Note that the leading OPE contribution to the sum of disconnected
contributions comes from perturbative contributions which are fourth order in
the light-quark masses. These contributions to $\r^{\rm EM}(s)$ are suppressed
by a factor of $(m_s^2-m_l^2)^2/(N_c s^2)$, the fourth order mass
dependence arising because two mass insertions are required in each of the
disconnected loops if the loop integral is to survive after the sum over
all of $u$, $d$ and $s$ running around the loop is performed.}
Based on the experimental observation that the $\r$ meson
spectrum and the $\o$ meson spectrum are nearly degenerate,\footnote{We observe that the
first three resonances are nearly degenerate, and have approximately equal width over mass
ratios (except the $\o(782)$, for which the width is restricted by phase space).} 
we will assume that, far enough above the narrow $\o(782)$ resonance, the duality violating part of the non-strange $I=0$ spectral function is degenerate in shape with that of the $I=1$ spectral function.  For the strange $I=0$ part we will use a parametrization as in Eq.~(\ref{ansatz}), 
but not assume that all parameters are the same as those for the non-strange part. 
Taking into account the relevant charge factors, we then arrive at the {\it ansatz}
\begin{equation}
\label{EMansatz}
\r_{\rm DV}^{\rm EM}(s)=\frac{5}{9}\,e^{-\d_1-\g_1 s}\sin(\a_1+\b_1 s)
+\frac{1}{9}\,e^{-\d_0-\g_0 s}\sin(\a_0+\b_0 s)\ .
\end{equation}
We emphasize that, while the framework of Ref.~\cite{BCGMP} provides strong arguments for
the use of such an {\it ansatz} in the $SU(3)$ chiral limit (in which $\d_0=\d_1$, \etc), additional assumptions are needed
in order to arrive at this form.   The factor $5/9$  has been chosen such that 
the expression $e^{-\d_1 - \g_1 s} \sin(\a_1 +\b_1 s)$ corresponds, in the isospin limit,  to the
duality violating $I=1$ contribution $\rho_{DV}^{I=1}(s)$ employed in the analysis of hadronic $\t$ decays in Ref.~\cite{alphas1,alphas2,alphas14}.
The factor $1/9$ is the square of the strange quark charge.   In this form, the $I=0$ and $I=1$ 
duality-violation parameters must become equal in the $SU(3)$ limit. 
Some shifts are, however, expected away from this limit, \eg, to take into account the fact that
the resonance peaks in the strange $I=0$ contributions are shifted to higher $s$.
 
Even the form~(\ref{EMansatz}) is not directly usable given the quality of the data we will be
working with, and more simplifications are needed.   First, we will take the $I=1$ parameters
$\d_1$, $\g_1$, $\a_1$ and $\b_1$ and their associated covariances from
the sum-rule analysis of hadronic $\t$-decay data reported in
Refs.~\cite{alphas2,alphas14}.    As we will see below, this strategy is reasonable since 
$\r_{\rm DV}^{I=1}(s)$, with parameters taken from the $\t$ analysis,
leads to an acceptable description of the $I=1$ component of the $R$-ratio 
data.   Furthermore, we will take $\b_0=\b_1$, as this parameter is
directly proportional to the asymptotic Regge slope, which we will assume not to be affected by $SU(3)$ flavor symmetry breaking.   Likewise, we will assume, as an approximation, $\g_0=\g_1$,\footnote{This corresponds to neglecting the difference between the widths of the $\r$ and $\o$ resonances and the, in general somewhat smaller, widths of the $\f$ resonances in the 
same mass region.} thus leaving us with only the two new free parameters $\d_0$ and $\a_0$.   

All these assumptions put significant limitations on our ability to study duality violations in the 
case of the EM vacuum polarization.   We emphasize however that, as we will see below, our main results for $\a_s$ will come from fits for which duality violations can be neglected; fits including duality violations will only serve as a consistency check on our central values
and provide us with a means of estimating the systematic uncertainty
resulting from neglecting these contributions.   In contrast to the case
of hadronic $\t$ decays, where data are limited to the region $s\le m_\t^2$, in the case
of $e^+e^-\to\mbox{hadrons}$ we can go to larger $s$, where duality violations turn out to be
less significant, as one would expect.

\section{\label{data} Data}
In this section, we discuss the experimental data for $R(s)$ employed in
the fits described in this paper.   Our data for $R(s)$ are taken from a new compilation, incorporating all available experimental results, presented first in Ref.~\cite{KNT18}, where this compilation 
was used for new determinations of the hadronic vacuum polarization contribution to 
the muon anomalous magnetic moment
and the QED coupling at the scale $M_Z$, $\a(M_Z^2)$.

\begin{figure}[t]
\vspace*{4ex}
\begin{center}
\includegraphics*[width=12cm]{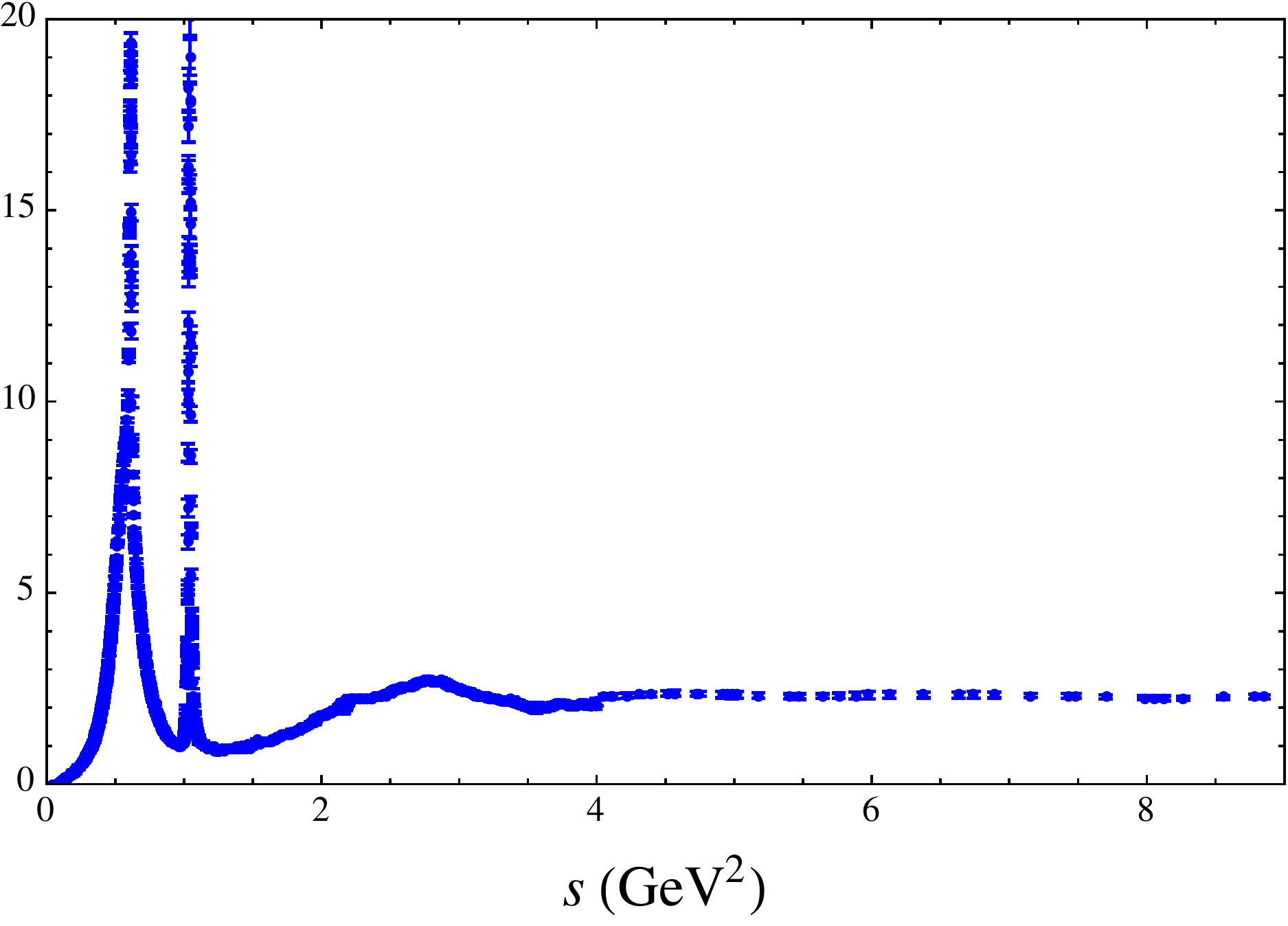}
\end{center}
\begin{quotation}
\floatcaption{Rratio}%
{{\it $R$-ratio data from Ref.~\cite{KNT18}, as a function of $s$, the hadronic invariant
squared mass.   The parton-model value in this 
region is $2$. See Fig.~\ref{Rblowup} for a comparison with perturbation theory.}}
\end{quotation}
\vspace*{-4ex}
\end{figure}

The data are shown in Fig.~\ref{Rratio}, where they are plotted against $s$, the square
of the center-of-mass energy for the process $e^+e^-\to\mbox{hadrons}(\g)$.   The plot is restricted to results on the interval from $s=0$ to 9~GeV$^2$, just below the charm threshold, which, as we will see below, is the region most relevant for our fits.  For more figures 
showing these data, we refer to Ref.~\cite{KNT18}.

\begin{figure}[t]
\vspace*{4ex}
\begin{center}
\includegraphics*[width=12cm]{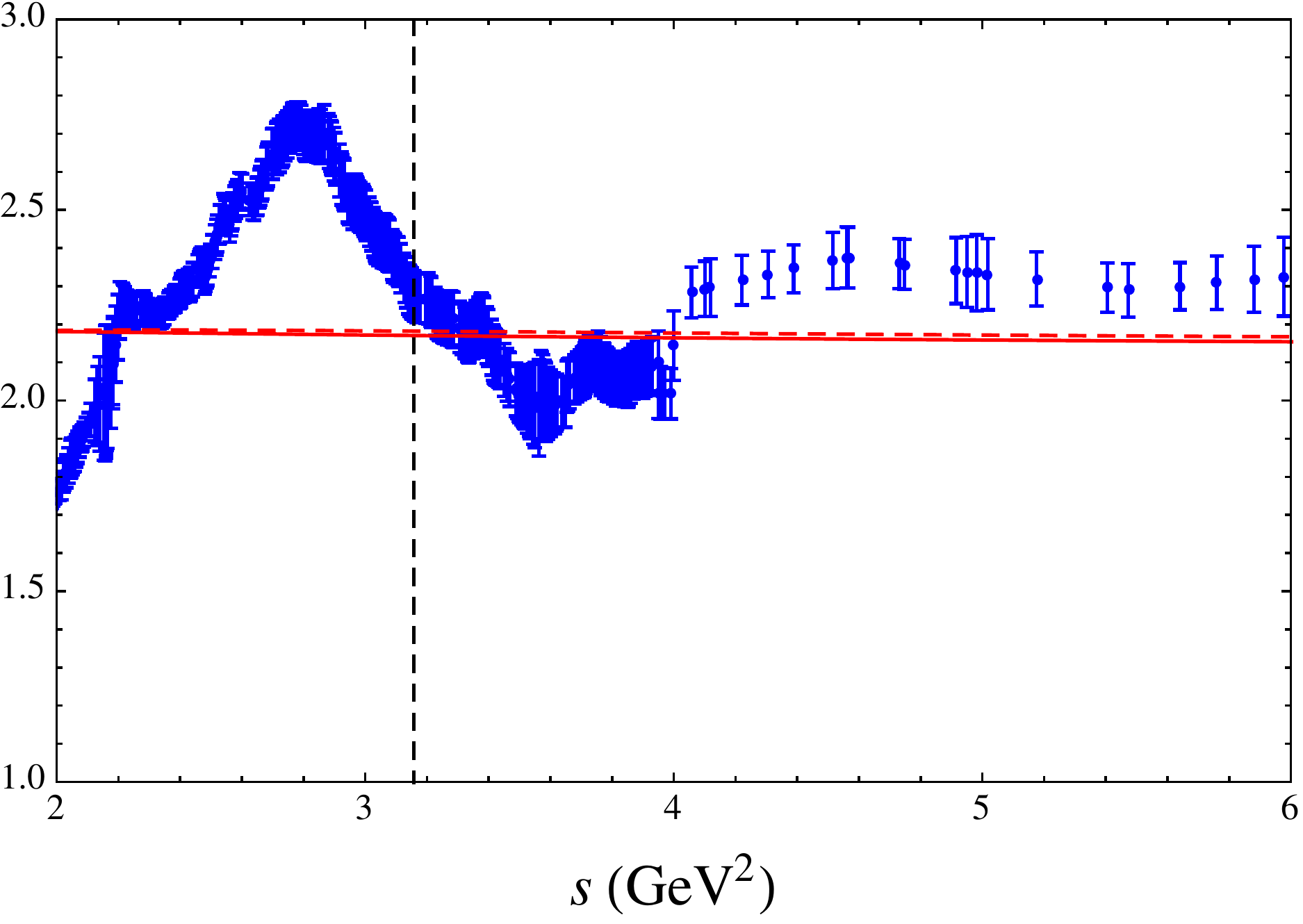}
\end{center}
\begin{quotation}
\floatcaption{Rblowup}%
{{\it A blow-up of the region $2\le s\le 6$~{\rm GeV}$^2$ in Fig.~\ref{Rratio}.
The red solid and red dashed lines show the results obtained
from perturbation theory with $\a_s(m_\t^2)=0.28$ and $\a_s(m_\t^2)=0.32$,
respectively.   The vertical dashed
line is $s=m_\t^2$.}}
\end{quotation}
\vspace*{-4ex}
\end{figure}

\subsection{\label{inclexcl} Inclusive {\textit{vs}.} exclusive data}
In Fig.~\ref{Rblowup}, we show a blow-up of Fig.~\ref{Rratio}, focussing on the region
$2\le s\le 6$~GeV$^2$.  The vertical axis range shown is centered on the parton-model value, 
$R=2$, for this region.

One difference that should be pointed out between the data set used here and that employed in Ref.~\cite{KNT18} is the choice concerning the data input for $R(s)$ at about $4$~GeV$^2$. Below this energy, the $R$-ratio is obtained as a sum over all exclusive hadronic channels. Results for each individual hadronic channel are obtained by combining the available data from many different experiments, where the combination procedure fully incorporates all available correlated uncertainties into the determination of the mean values and uncertainties of the combined cross section. Above about $4$~GeV$^2$, $R(s)$ is instead obtained from the available measured inclusive data (all hadronic channels) using the same procedure to combine the inclusive data from different experiments as with the exclusive channels.\footnote{Below about $4$~GeV$^2$, it becomes increasingly difficult to experimentally measure the inclusive $R$-ratio and requires a detailed understanding of the experimental efficiencies for exclusive states which contribute. Older inclusive measurements do exist slightly below $4$~GeV$^2$ (see the discussions in~Refs.~\cite{KNT18,HMNT03,HMNT06,HLMNT11} concerning these data). However, these data are imprecise and of poor quality, making them impractical for use in the determination of $R(s)$. In addition, very few of the exclusive states contributing to the hadronic $R$-ratio have been measured above $4$~GeV$^2$. For details concerning all combined experimental data, we refer to Ref.~\cite{KNT18}.} 
The inclusive data combination extends only down to $s$ around $3.39$ GeV$^2$. Moreover, in the lower part of this region, few such data points are available. In principle, one could use either the sum of exclusive states or the inclusive data combination in the range $3.39\leq s\leq 4$~GeV$^2$. 
In Ref.~\cite{KNT18}, the choice was made to transition from the sum of exclusive states to the inclusive data at $s\sim3.75$~GeV$^2$. However, in the region of overlap, the results obtained by summing exclusive data are more precise.
In this work, we have chosen to retain the full information from the sum of exclusive channels up to $s=4$~GeV$^2$, for reasons which we will now discuss in more detail.

The determination of $\a_s$ from electroproduction in this paper is very similar to that from hadronic
$\t$ decays, but has the advantage that the experimental spectral data are kinematically unconstrained and hence are available above $s= m_\t^2$.
We would thus like to use the full range of available $R$-ratio data, up to at least the 
charm threshold at $s\sim 9$~GeV$^2$.   However, as we will see below, the errors on the data in the inclusive region, $s>4$~GeV$^2$, are too large to allow for a precision determination of $\a_s$ in
which these data play a major role.  

\begin{figure}[t]
\vspace*{4ex}
\begin{center}
\includegraphics*[width=7.3cm]{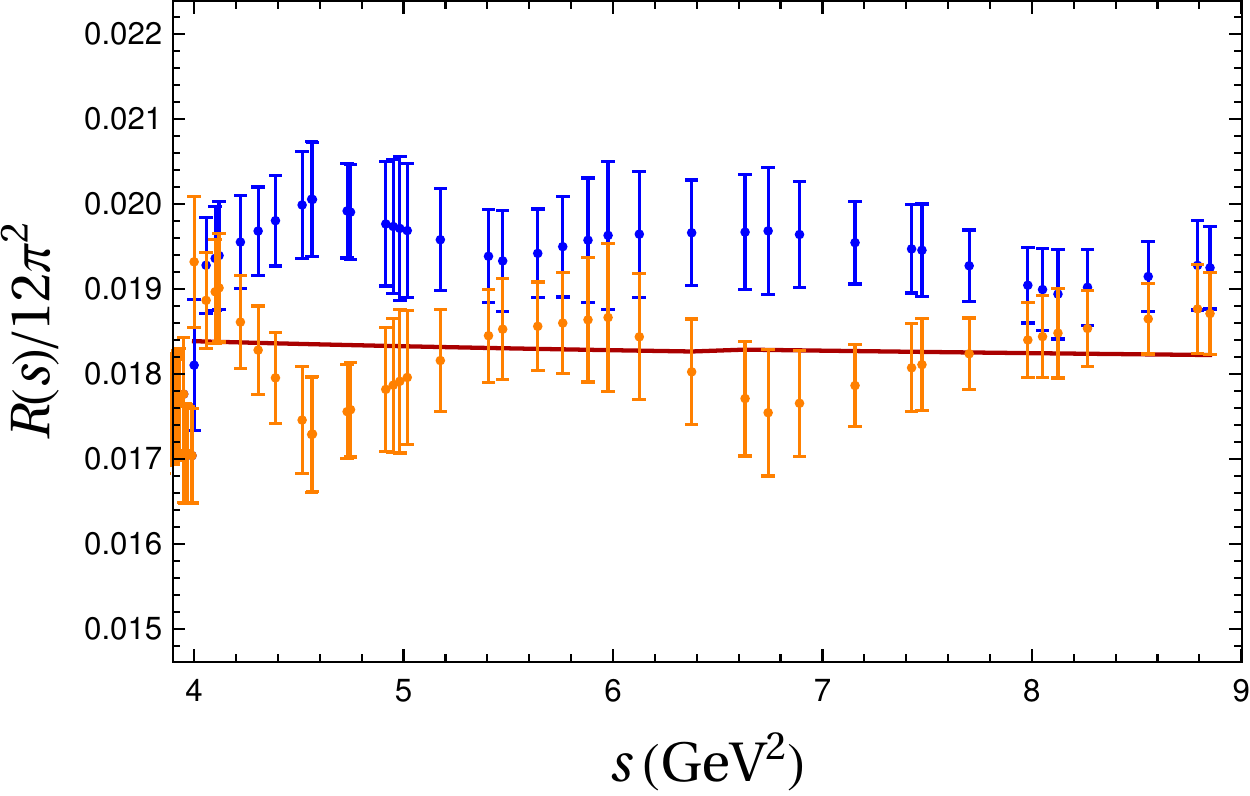}
\hspace{0.2cm}
\includegraphics*[width=7.3cm]{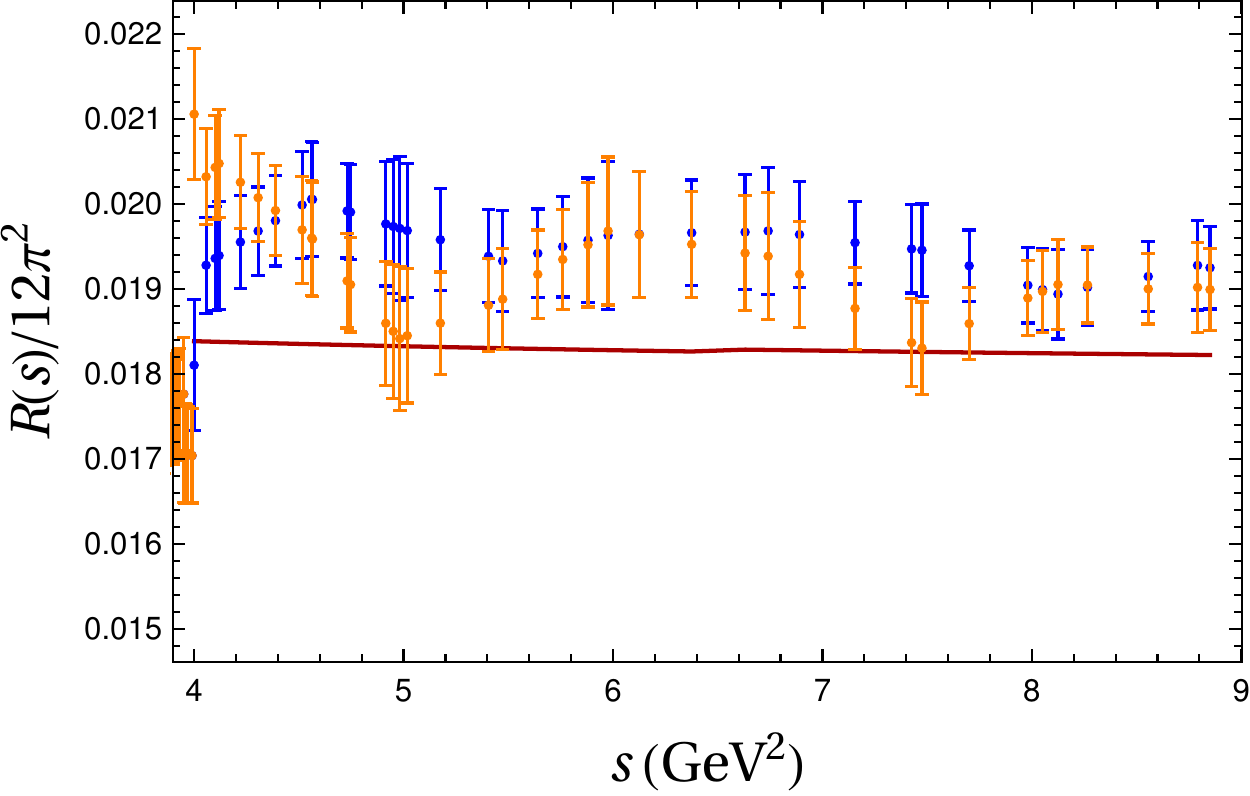}
\end{center}
\begin{quotation}
\floatcaption{mockdata}%
{{\it Two mock data sets generated with the covariance matrix of the real data, in the inclusive
region, $4<s<9$~{\rm GeV}$^2$.  The blue data points are the actual data, the orange data points the mock data.   The curve shows perturbation theory with
$\a_s(m_\t^2)=0.3$.}}
\end{quotation}
\vspace*{-4ex}
\end{figure}

In Fig.~\ref{Rblowup}, we also show the theoretical prediction for $R(s)$ from five-loop perturbation
theory (including the six-loop estimate $c_{51}=283$); the solid red curve corresponds to
$\a_s(m_\t^2)=0.28$, the dashed red curve to $\a_s(m_\t^2)=0.32$.  The data in the inclusive region $s>4$~GeV$^2$ all lie above the perturbative prediction.   Contrary to what one might naively conclude, however, this does not imply that the data are inconsistent with the expectations of perturbation theory, but rather reflects the size of the errors and the influence of the strong correlations present in the inclusive data. In order to investigate this question, we took perturbation theory, with $\a_s(m_\t^2)=0.3$
and used the actual data covariance matrix to generate several mock data sets, drawn from the same
distribution as the experimental data, but with central values determined by perturbation theory,
\ie, the $D=0$ part of Eq.~(\ref{OPE}).\footnote{We note that the experimental covariance matrix is not singular in the region $4<s<9$~{\rm GeV}$^2$.}

Generating a small number of mock data sets yielded the two sets
shown in Fig.~\ref{mockdata}.   The left-hand panel shows a mock
data set that, by eye, is perfectly consistent with perturbation theory, while the right-hand
panel shows a set very similar to the actual experimental data.   These examples
demonstrate that there is no
inconsistency between the data and perturbation theory.  Instead, the apparent 
discrepancy between the actual data and perturbation theory is consistent with a statistical fluctuation
caused by the non-trivial influence of the strong correlations in this region.  Of course, this is 
reassuring.  However, it also implies that the existing
inclusive $R(s)$ data set places only weak constraints on perturbation 
theory.   This is unfortunate, as perturbation theory becomes more reliable at larger $s$.
More precise data would be needed in this region to make
an impact on the determination of $\a_s$ from electroproduction data.   The upshot is that the
precision of our electroproduction-based
determination of $\a_s$ will be almost entirely driven by data from the exclusive region $s\leq 4$~GeV$^2$.

\subsection{\label{phiprimepeak} Nature of the peak at $s\sim 2.8$~GeV$^2$}
Next, let us consider the data in the region $2\le s\le 4$~GeV$^2$.   First,
even though the determination of $\a_s$ benefits primarily from the region 
 $s\le 4$~GeV$^2$, we note that this allows us to work at scales significantly 
higher than the maximum, $s=m_\t^2=3.157$~GeV$^2$, accessible in hadronic 
$\tau$ decays (shown as the vertical dashed line in Fig.~\ref{Rblowup}).
It is, however, clear from Fig.~\ref{Rblowup} that non-negligible duality 
violations remain present in the spectral function in this 
region.\footnote{Apparent faint oscillations in the inclusive data above $4$~GeV$^2$ are,
in contrast, not statistically significant.}
The question that remains is, of course, how much they affect the determination of $\a_s$.

\begin{figure}[t]
\vspace*{4ex}
\begin{center}
\includegraphics*[width=12cm]{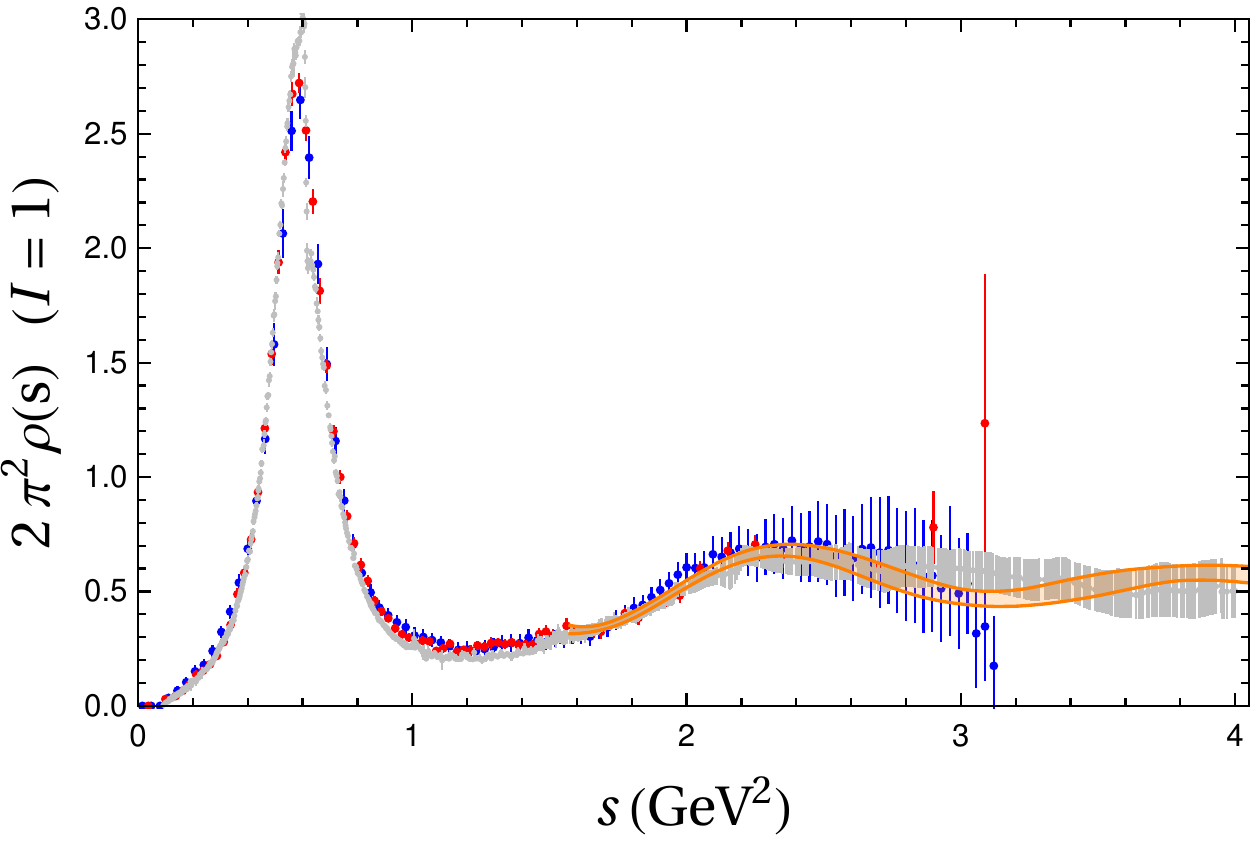}
\end{center}
\begin{quotation}
\floatcaption{Rbreakdown}%
{{\it The $I=1$ spectral function as a function of $s$.   Shown are the OPAL $\t$ data from Ref.~\cite{OPAL}
in blue, the ALEPH $\t$ data from Ref.~\cite{ALEPH13} in red, and the $I=1$ $R$-ratio data in gray.
All data have been normalized such that the parton-model version of the spectral function 
(not shown in the figure) would be a horizontal line at $2\p^2 \r(s)=1/2$. The orange band shows the fit to ALEPH data from Ref.~\cite{alphas14} described in Sec.~\ref{tests},
for $s$ extending down to the lowest value, $1.575$~{\rm GeV}$^2$,
for which the duality-violation \ansatz\ was employed in that fit.}}
\end{quotation}
\vspace*{-4ex}
\end{figure}

To understand this region in more detail, we attempted a separation of the $R$-ratio data
into $I=0$ and $I=1$ parts.   The result is shown in Fig.~\ref{Rbreakdown}.
This separation follows closely the strategy employed by ALEPH \cite{ALEPH} and 
OPAL \cite{OPAL} in separating vector and axial vector contributions 
to the non-strange hadronic $\tau$ decay distribution. 

In the electroproduction case the separation relies on the observation that
the isovector current $J_\mu^3$ is $G$-parity even and the isoscalar
current $J_\mu^8$ $G$-parity odd. Up to isospin-breaking corrections,
which should be safely small away from the low-$s$ regions near the narrow
$\omega$ and $\phi$ resonances, where such corrections can be locally
enhanced by resonance interference effects, $G$-parity can thus be used to 
uniquely assign the contributions of exclusive modes with well-defined 
$G$-parity to either the $I=0$ or $I=1$ channel. A significant fraction
of the exclusive modes contributing to $R(s)$ in the region
below $s=4$~GeV$^2$, in fact, have definite $G$-parity. States consisting
of an even (odd) number of pions only, for example, can be uniquely assigned 
to the $I=1$ ($I=0$) channel. Exclusive states involving, in addition
to some number of pions, also a $G$-parity even $\eta$ or $G$-parity odd
$\omega$ or $\phi$ are, similarly, uniquely assignable using $G$-parity.
States for which such a unique $G$-parity assignment is not possible
are those containing a $K\bar{K}$ pair not identifiable as coming from
the $\phi$ resonance. Among such states, additional information is
available only for $K\bar{K}\pi$, where BaBar \cite{babarkkpi}
observed a dominance by $K^*K$ below $s\simeq 4$~GeV$^2$ and performed
a Dalitz plot analysis to separate the $I=0$ and $1$ components of the
$K^*K$ cross-section. We take advantage of these results. Contributions
from modes lacking a unique $G$-parity assignment, and for which no 
additional information on the isospin separation is available, are 
treated in a maximally conservative manner by assigning to each of the 
$I=0$ and $1$ channels $(50\pm 50)\%$ of the sum of these contributions.
The results of this separation exercise are shown in Fig.~\ref{Rbreakdown}, for $I=1$.

This figure shows the data for the $I=1$ part of the EM spectral function in gray.
It shows that these data are in good agreement with data for 
the corresponding spectral functions obtained from hadronic $\t$ decays by OPAL \cite{OPAL},
shown in blue, and ALEPH \cite{ALEPH13}, shown in red.   The orange band shows the
results of one of the fits of Ref.~\cite{alphas14} to the ALEPH $\t$ data, 
starting from $s$ where the previous analysis suggests the asymptotic
duality-violation \ansatz~(\ref{ansatz}) is valid (described in more
detail and employed in Sec.~\ref{tests} below).   To the extent that the $\t$-based data and
the $I=1$ part of the EM data agree, it is clear that this fit also provides a reasonable
representation of the $I=1$ EM data, although the figure suggests that the $I=1$ EM data
might prefer a somewhat smaller value of $\b_1$ (with accompanying adjustments in the other
$I=1$ parameters).  

\section{\label{analysis} Analysis}
In this section, we will present our main analysis, employing the sum rule~(\ref{cauchy}) with
weights~(\ref{weights}).   At first, we will ignore duality violations, while retaining all relevant terms in the OPE~(\ref{OPEdef}), with the assumptions detailed in Sec.~\ref{OPE}.   To perform these fits, we
need the integrated data, as a function of $s_0$, \ie, the integrals $I^{(w)}(s_0)$ of Eq.~(\ref{cauchy}).
We perform the fits of these integrals as
a function of $s_0$, ranging from a value $s_0^{\rm min}$ between $2.5$ and $3.8$~GeV$^2$
to $s_0^{\rm max}=4$ GeV$^2$,
with the separations of adjacent $s_0$ as close as possible to $\Delta s_0=0.05$ GeV$^2$.    In some cases, it turns out that the integrated data are too strongly correlated to obtain
good fits (as measured by their $p$-value), in which case we enlarge the spacing to $\D s_0\approx 0.1$~GeV$^2$.   We will refer to
this procedure as ``thinning'' by a factor 2.   For more details on the use of thinning, we refer to 
Sec.~\ref{fits}.   It should be noted that even using the spacing $\D s_0\approx 0.05$~GeV$^2$ corresponds
to a thinning of the data, because throughout the spectrum below 4~GeV$^2$, the binning of the data
is much finer than $0.05$~GeV$^2$.

The central values for the weighted spectral integrals $I^{(w_i)}(s_0)$, $i=0,2,3,4$, on the left-hand side
of Eq.~(\ref{cauchy}) are obtained from the data using the trapezoidal rule.\footnote{We checked that using a different method, such as a histogram rule, makes no
significant difference.}   Despite the fact that the integrated data, \ie, the moments $I^{(w_i)}(s_0)$,
are strongly correlated between different values of $s_0$, we find that these integrated data allow
us to perform fully correlated fits, on the interval $2.5$~GeV$^2\le s_0\le 4.0$~GeV$^2$.    It is thus the results of these correlated fits that we present in this paper.

\begin{table}[t!]
\begin{center}
\begin{tabular}{|c|c|c|c|c|}
\hline
$s_0^{\rm min}$ (GeV$^2$) & \# dofs & $\c^2$  & $p$-value & $\alpha_s$  \\
\hline
\hline
3.00 & 20 & 76.5 & 2$\times 10^{-8}$ & 0.233(13)\\
3.15 & 17 & 34.6 & 0.007 & 0.275(13)\\
3.25 & 15 & 27.8 & 0.02 & 0.287(14)\\
3.00 & 10* & 53.3 & 7$\times 10^{-8}$ & 0.236(13)\\
3.15 & 8* & 16.0 & 0.043 & 0.279(13)\\
\hline
3.25 & 7* & 9.33 & 0.23 & 0.292(14)\\
3.35 & 13 & 19.0 & 0.12 & 0.297(14)\\
3.45 & 11 & 14.9 & 0.19 & 0.304(14)\\
3.55 & 9 & 14.2 & 0.12 & 0.302(15)\\
3.60 & 8 & 10.8 & 0.21 & 0.304(15)\\
3.70 & 6 & 7.21 & 0.30 & 0.296(16)\\
3.80 & 4 & 6.98 & 0.14 & 0.298(17)\\
\hline
\hline
3.00 & 20 & 76.4 & 2$\times 10^{-8}$ & 0.236(14)\\
3.15 & 17 & 34.6 & 0.007 & 0.282(15)\\
3.25 & 15 & 28.0 & 0.02 & 0.295(16)\\
3.00 & 10* & 53.2 & 7$\times 10^{-8}$ & 0.239(14)\\
3.15 & 8* & 16.0 & 0.04 & 0.287(16)\\
\hline
3.25 & 7* & 9.64 & 0.21 & 0.301(17)\\
3.35 & 13 & 19.6 & 0.11 & 0.306(17)\\
3.45 & 11 & 15.7 & 0.15 & 0.314(17)\\
3.55 & 9 & 14.9 & 0.09 & 0.311(18)\\
3.60 & 8 & 11.6 & 0.17 & 0.313(18)\\
3.70 & 6 & 7.65 & 0.27 &  0.305(18)\\
3.80 & 4 & 7.46 & 0.11 &  0.306(20)\\
\hline
\end{tabular}
\end{center}
\floatcaption{tab1}{\it Fits to $I^{(w_0)}(s_0)$
from $s_0=s_0^{\rm min}$ to $s_0=s_0^{\rm max}=4$~{\rm GeV}$^2$. FOPT results
are shown above the double line, CIPT below.  
Fits below the single horizontal lines are used in the average of Eq.~(\ref{alphasw0});
those marked with an asterisk are thinned by a factor 2.}
\end{table}%

\subsection{\label{fits} Fits}
In Table~\ref{tab1} we show the results for fits using the weight $w_0=1$, for a range of choices of $s_0^{\rm min}$.     As the weight $w_0$ is unpinched, the FESR for this weight is the most 
susceptible to possible non-negligible duality-violating effects.
The first column gives the values of $s_0^{\rm min}$ employed, the second column
the number of degrees of freedom in the fit, \ie, the number of $s_0$ values between $s_0^{\rm min}$ and $s_0^{\rm max}$ minus the number of parameters in the fit.   The third column gives the minimum of $\c^2$ found in the fit, the fourth
column the corresponding $p$-value, and the final column the value of $\a_s$ obtained in the fit.   Results above (below) the double horizontal line are obtained using FOPT (CIPT).    

It is obvious that the fit quality increases strongly with increasing
$s_0^{\rm min}$, as does the value of $\a_s$, with the latter leveling off when the fits become good, 
and peaking at $s_0^{\rm min}\approx 3.45$~GeV$^2$, after which it decreases somewhat.   
We find that, for 
$s_0^{\rm min}=3.25$~GeV$^2$, the quality of the fits improves significantly if we thin out the data by a factor
2 (\ie, use $\D s_0=0.1$~GeV$^2$), as shown in Table~\ref{tab1}:  the $p$-values increase, while the fit parameters remain stable.   For  $s_0^{\rm min}<3.25$~GeV$^2$, there is no clear improvement from thinning out, and $p$-values are bad or marginal.   (We will return to fits with these values of
$s_0^{\rm min}$ in Sec.~\ref{tests} below.)   For higher values of $s_0^{\rm min}$,
the fits are already good, and do not improve significantly with thinning.
By $p$-values, the fits with
$s_0^{\rm min}$ ranging from $3.25$ to $3.80$~GeV$^2$ are preferred; in the table, they are the fits 
below the single horizontal lines.  Averaging these values of $\a_s$ yields the estimates
\begin{equation}
\label{alphasw0}
\a_s(m_\t^2)|_{w_0}=\left\{\begin{array}{ll}
0.299(15)(6)&\qquad\mbox{(FOPT)}\ ,\cr
0.308(18)(6)&\qquad\mbox{(CIPT)\ .}
\end{array}\right.
\end{equation}
These values were obtained by a simple average; while one can devise various weighted
averages, they all yield very similar results.   The first error is the average fit error, the second
half the difference between the lowest and highest value entering the average.   As Table~\ref{tab1}
shows, the variation in the values of $\a_s$ as a function of $s_0^{\rm min}$ is in fact smaller
than the average fit error of $\pm 0.015$ and $\pm 0.018$, for FOPT and CIPT, respectively, 
and might also be statistical in nature.   However,
since these values of $\a_s$ are highly correlated, it is likely that there is a systematic
component as well.   Hence, we choose to be conservative, and show the second error
as a separate error.

\begin{table}[t!]
\begin{center}
\begin{tabular}{|c|c|c|c|c|c|}
\hline
$s_0^{\rm min}$ (GeV$^2$) & \# dofs & $\c^2$ & $p$-value & $\alpha_s$ & $C_6$ in GeV$^6$ \\
\hline
\hline
3.00 & 19 & 53.4 & 0.00004 & 0.239(13) & -0.0027(13) \\
3.15 & 16 & 25.1 & 0.07 & 0.278(14) & 0.0033(19) \\
3.00 & 9* & 38.0 & 0.00002 & 0.253(13) & -0.0011(15) \\
3.15 & 7* & 13.6 & 0.06 & 0.287(14) & 0.0049(21) \\
\hline
3.25 & 14 & 17.3 & 0.24 & 0.292(14) & 0.0062(23) \\
3.35 & 12 & 13.6 & 0.33 & 0.298(15) & 0.0078(26) \\
3.45 & 10 & 10.3 & 0.42 & 0.305(15) & 0.0097(27) \\
3.50 & 8 & 9.45 & 0.31 & 0.302(16) & 0.0088(30) \\
3.60 & 7 & 9.45 & 0.22 & 0.302(16) & 0.0088(31) \\
3.70 & 5 & 5.32 & 0.38 & 0.293(16) & 0.0057(34) \\
3.80 & 3 & 5.14 & 0.16 & 0.296(18) & 0.0064(38) \\
\hline
\hline
3.00 & 19 & 53.3 & 0.00004 & 0.242(14) & -0.0029(13) \\
3.15 & 16 & 25.2 & 0.07 & 0.284(15) & 0.0026(17) \\
3.00 & 9* & 37.9 & 0.00002 & 0.257(14) & -0.0013(14) \\
3.15 & 7* & 13.8 & 0.06 & 0.294(16) & 0.0040(18) \\
\hline
3.25 & 14 & 17.6 & 0.23 & 0.298(16) & 0.0051(20) \\
3.35 & 12 & 14.0 & 0.30 & 0.306(17) & 0.0065(22) \\
3.45 & 10 & 10.8 & 0.37 & 0.313(17) & 0.0081(23) \\
3.55 & 8 & 9.90 & 0.32 & 0.309(18) & 0.0073(25) \\
3.60 & 7 & 9.90 & 0.19 & 0.309(18) & 0.0073(26) \\
3.70 & 5 & 5.57 & 0.35 & 0.300(18) & 0.0045(29) \\
3.80 & 3 & 5.42 & 0.14 & 0.302(19) & 0.0050(32) \\
\hline
\end{tabular}
\end{center}
\floatcaption{tab2}{\it Fits to $I^{(w_2)}(s_0)$
from $s_0=s_0^{\rm min}$ to $s_0=s_0^{\rm max}=4$~{\rm GeV}$^2$. FOPT results
are shown above the double line, CIPT below.
Fits below the single horizontal lines are used in the average of Eq.~(\ref{alphasw2});
those marked with an asterisk are thinned by a factor 2.}
\end{table}%

\begin{table}[h!]
\begin{center}
\begin{tabular}{|c|c|c|c|c|c|c|}
\hline
$s_0^{\rm min}$ (GeV$^2$) & \# dofs & $\c^2$ & $p$-value & $\alpha_s$ & $C_6$ in GeV$^6$ 
& $C_8$ in GeV$^8$\\
\hline
\hline
3.15 & 15 & 44.8 & 0.00008& 0.276(15) & 0.0027(20) & -0.0184(51) \\
3.25 & 13 & 31.9 & 0.003 & 0.292(15) & 0.0059(23) & -0.0278(61) \\
3.35 & 11 & 26.0 & 0.006 & 0.296(15) & 0.0068(25) & -0.0305(67) \\
3.15 & 6* & 9.79 & 0.13& 0.293(15) & 0.0055(22) & -0.0261(57) \\
\hline
3.25 & 5* & 7.60 & 0.18 & 0.299(15) & 0.0070(25) & -0.0307(65) \\
3.35 & 4* & 5.62 & 0.23 & 0.305(16) & 0.0084(27) & -0.0353(73) \\
3.45 & 9 & 12.9 & 0.17 & 0.303(16) & 0.0085(27) & -0.0360(75) \\
3.55 & 7 & 11.6 & 0.11 & 0.301(16) & 0.0081(29) & -0.0346(83)  \\
3.60 & 6 & 11.1 & 0.09 & 0.298(17) & 0.0071(32) & -0.0311(95)  \\
3.70 & 4 & 5.68 & 0.22 & 0.292(18) & 0.0049(35) & -0.023(11) \\
3.80 & 2 & 2.31 & 0.32 & 0.289(19) & 0.0036(39) & -0.019(12) \\
\hline
\hline
3.15 & 15 & 44.9 & 0.00008 & 0.279(13) & 0.0022(15) & -0.0177(41) \\
3.25 & 13 & 32.2 & 0.002 & 0.297(16) & 0.0051(20) & -0.0266(56) \\
3.35 & 11 & 26.4 & 0.006 & 0.301(17) & 0.0059(22) & -0.0290(64) \\
3.15 & 6* & 9.94 & 0.13 & 0.298(16) & 0.0047(19) & -0.0250(54) \\
\hline
3.25 & 5* & 7.86 & 0.16 & 0.305(17) & 0.0061(22) & -0.0293(62) \\
3.35 & 4* & 5.97 & 0.20 & 0.310(17) & 0.0074(24) & -0.0336(70) \\
3.45 & 9 & 13.3 & 0.15 & 0.308(17) & 0.0075(24) & -0.0342(72) \\
3.55 & 7 & 12.0 & 0.10 & 0.306(18) & 0.0070(26) & -0.0329(79)  \\
3.60 & 6 & 11.4 & 0.08 & 0.303(18) & 0.0061(29) & -0.0294(91)  \\
3.70 & 4 & 5.87 & 0.21 & 0.297(19) & 0.0040(31) & -0.022(10) \\
3.80 & 2 & 2.45 & 0.29 & 0.293(20) & 0.0028(35) & -0.017(12) \\
\hline
\end{tabular}
\end{center}
\floatcaption{tab3}{\it Fits to $I^{(w_3)}(s_0)$
from $s_0=s_0^{\rm min}$ to $s_0=s_0^{\rm max}=4$~{\rm GeV}$^2$. FOPT results
are shown above the double line, CIPT below.
Fits below the single horizontal lines are used in the average of Eq.~(\ref{alphasw3});
those marked with an asterisk are thinned by a factor 2.}
\end{table}%

\begin{table}[h!]
\begin{center}
\begin{tabular}{|c|c|c|c|c|c|c|}
\hline
$s_0^{\rm min}$ (GeV$^2$) & \# dofs & $\c^2$ & $p$-value  & $\alpha_s$ & $C_6$ in GeV$^6$ 
& $C_{10}$ in GeV$^{10}$\\
\hline
\hline
3.15 & 15 & 45.0 & 0.00008 & 0.275(15) & 0.0027(20) & 0.079(14) \\
3.25 & 13 & 32.0 & 0.002 & 0.292(15) & 0.0060(24) & 0.107(17) \\
3.35 & 11 & 26.0 & 0.006 & 0.296(15) & 0.0069(25) & 0.115(19) \\
3.15 & 6* & 9.76 & 0.14 & 0.292(15) & 0.0056(22) & 0.101(16) \\
\hline
3.25 & 5* & 7.55 & 0.18 & 0.299(15) & 0.0071(25) & 0.115(18) \\
3.35 & 4* & 5.59 & 0.23 & 0.304(15) & 0.0086(27) & 0.130(21) \\
3.45 & 9 & 12.9 & 0.17 & 0.302(16) & 0.0087(28) & 0.133(22) \\
3.55 & 7 & 11.6 & 0.11 & 0.300(16) & 0.0082(30) & 0.129(25) \\
3.60 & 6 & 11.0 & 0.09 & 0.297(17) & 0.0072(32) & 0.117(30) \\
3.70 & 4 & 5.69 & 0.22 & 0.292(18) & 0.0050(35) & 0.089(34) \\
3.80 & 2 & 2.30 & 0.32 & 0.288(19) & 0.0037(39) & 0.072(40) \\
\hline
\hline
3.15 & 15 & 45.2 & 0.00007 & 0.279(16) & 0.0022(17) & 0.077(123) \\
3.25 & 13 & 32.3 & 0.002 & 0.297(13) & 0.0051(15) & 0.104(12) \\
3.35 & 11 & 26.4 & 0.006 & 0.301(17) & 0.0059(22) & 0.112(18) \\
3.15 & 6* & 9.92 & 0.13 & 0.298(16) & 0.0047(19) & 0.098(15) \\
\hline
3.25 & 5* & 7.82 & 0.17 & 0.305(17) & 0.0061(22) & 0.112(18) \\
3.35 & 4* & 5.96 & 0.20 & 0.310(17) & 0.0074(24) & 0.126(20) \\
3.45 & 9 & 13.3 & 0.15 & 0.308(17) & 0.0075(24) & 0.129(21) \\
3.55 & 7 & 12.0 & 0.10 & 0.306(18) & 0.0071(26) & 0.124(24) \\
3.60 & 6 & 11.4 & 0.08 & 0.303(18) & 0.0061(29) & 0.112(29) \\
3.70 & 4 & 5.90 & 0.21 & 0.297(19) & 0.0040(31) & 0.084(33) \\
3.80 & 2 & 2.44 & 0.30 & 0.293(20) & 0.0028(35) & 0.067(39) \\
\hline
\end{tabular}
\end{center}
\floatcaption{tab4}{\it Fits to $I^{(w_4)}(s_0)$
from $s_0=s_0^{\rm min}$ to $s_0=s_0^{\rm max}=4$~{\rm GeV}$^2$. FOPT results
are shown above the double line, CIPT below.
Fits below the single horizontal lines are used in the average of Eq.~(\ref{alphasw4});
those marked with an asterisk are thinned by a factor 2.}
\end{table}%

Before we
discuss further the results of the fits shown in Table~\ref{tab1}, we present the results from fits employing the
other three weights, $w_{2,3,4}$ of Eq.~(\ref{weights}).   They are collected in Tables~\ref{tab2} to
\ref{tab4}.    Table~\ref{tab2} shows good $p$-values for 
$s_0^{\rm min}$ between $3.25$ and $3.80$~GeV$^2$; thinning does not appear to improve the fit for $s_0^{\rm min}=3.15$~GeV$^2$.
Taking the average of the fits with $s_0^{\rm min}$ between $3.25$ and $3.80$~GeV$^2$ yields
\begin{equation}
\label{alphasw2}
\a_s(m_\t^2)|_{w_2}=\left\{\begin{array}{ll}
0.298(16)(6)& \qquad\mbox{(FOPT)}\ ,\cr
0.305(18)(7)& \qquad\mbox{(CIPT)}\ .
\end{array}\right.
\end{equation}

For fits with the weights $w_3$ and $w_4$ we find that, for lower values of
$s_0^{\rm min}$, the quality of the fits improves significantly if we thin out the data by a factor
2 (\ie, use $\D s_0=0.1$~GeV$^2$), as shown in Tables~\ref{tab3}
and~\ref{tab4}:  the $p$-values increase, while, at least for $s_0^{\rm min}=3.25$
and $3.35$~GeV$^2$, the fit parameters remain stable.   Also the fit with $s_0^{\rm min}=3.15$~GeV$^2$ has a good $p$-value after thinning, but parameter values
are not stable, \seef\ Table~\ref{tab3}.\footnote{Fits thinned by a factor 3 (\ie, using $\D s_0=0.15$~GeV$^2$) with $s_0^{\rm min}=3.15$~GeV$^2$ cause the $p$-values to decrease to about $0.05$, but
yield stable fit parameters in comparison with the fit with $\D s_0=0.1$~GeV$^2$.  
One could, thus, also consider including the results of the thinned fits with $s_0^{\rm min}=3.15$~GeV$^2$ in the average. Since this turns out not to alter the average reported in Eq.~(\ref{alphasw3}) at the level of accuracy reported there, we choose to average here over the same set of $s_0^{\rm min}$ used in arriving at the $w_2$ average in Eq.~(\ref{alphasw2}). The same comments apply to the 
$w_4$ average reported in Eq.~(\ref{alphasw4}).} 
For higher values of $s_0^{\rm min}$,
the fits are already good, and do not improve significantly with thinning.

Table~\ref{tab3} shows good $p$-values for 
$s_0^{\rm min}$ between $3.25$ and $3.80$~GeV$^2$ if for $s_0^{\rm min}=3.25$ and $3.35$~GeV$^2$ we take the thinned fits; taking the average yields
\begin{equation}
\label{alphasw3}
\a_s(m_\t^2)|_{w_3}=\left\{\begin{array}{ll}
0.298(16)(8)& \qquad\mbox{(FOPT)}\ ,\cr
0.303(18)(8)& \qquad\mbox{(CIPT)}\ .
\end{array}\right.
\end{equation}
We note that the weight for which we report results in Table~\ref{tab4} just trades $C_8$
for $C_{10}$, and thus does not increase the number of parameters in the fits.   It shows
good $p$-values for $s_0^{\rm min}$ between $3.45$ and $3.80$~GeV$^2$ and for
$s_0^{\rm min}=3.25$ and $3.35$~GeV$^2$ if we thin as for $w_3$; taking the average yields
\begin{equation}
\label{alphasw4}
\a_s(m_\t^2)|_{w_4}=\left\{\begin{array}{ll}
0.297(16)(8)& \qquad\mbox{(FOPT)}\ ,\cr
0.303(18)(8)& \qquad\mbox{(CIPT)}\ .
\end{array}\right.
\end{equation}
In Fig.~\ref{fitplots} we show the fits for the lowest $s_0^{\rm min}$ value used in the averages reported
in Eqs.~(\ref{alphasw0}) to~(\ref{alphasw4}).    Other fits show equally good visual matches between 
data and fit curves.   The oscillatory behavior as a function of  $s_0$ seen in the data in the upper left panel of
Fig.~\ref{fitplots} is what one typically expects to see when integrated duality violations are not entirely negligible. Such residual duality violations are expected to be most visible for
the unpinched weight $w_0$.   The absence of oscillatory behavior in the other panels is
consistent with the suppression of duality violations by the pinching of the other weights.

The fit qualities ($p$-values) improve going from weight $w_0$
to weight $w_2$, especially for lower values of $s_0^{\rm min}$, as can be
seen by comparing corresponding fits in Tables~\ref{tab1} and \ref{tab2}.   This provides 
additional evidence that pinching indeed
suppresses duality violations (whether they are asymptotic, in the sense of being
described by Eq.~(\ref{EMansatz}), or not).   However, this improvement does not appear to persist with
more pinching, as can be seen in Tables~\ref{tab3} and \ref{tab4}.  There are several possible
reasons for this.    

One of these is that the theoretical model underlying the fits with weights
$w_3$ and $w_4$ may be less good than the one underlying the fit with weight $w_2$.
The higher-degree weights employed in these fits probe higher orders in the
OPE, and it is possible that with these higher-$D$ terms we enter the region (at these values of $s_0$) where the OPE converges less well.   An indication of this is that, for $s_0$ values
in the range $3$ to $4$~GeV$^2$, the $D=8$ and $D=10$ terms are of about the same
size as the $D=6$ term, if we employ the values for $C_{6,8,10}$ reported in these tables,
in the $s_0^{\rm min}$ range with good $p$-values.\footnote{It is also worth
noting that, from the results in Table~\ref{tab4}, the central $C_{10}$ value is large, and
lies many $\s$ from zero.   
Using the effective condensates from Tables~\ref{tab3} and \ref{tab4}, it is also easily
shown that the assumption made in a number of $\t$-based analyses that
integrated $D=10$ and higher contributions can be neglected, relative to
integrated lower dimension non-perturbative contributions, for $s_0$ as
large as $m_\t^2$ would fail quite badly for the analogous EM case considered
here.}   A possible interpretation is that use of the
weight $w_2$ provides an optimal balance between suppression of duality violations
(because of its zero at $s=s_0$), and the convergence properties of the OPE, in this
range.   We note that the $D=6$ contribution is always very small compared to the $D=0$
(\ie, purely perturbative) term.  

Another possibility is statistical in nature.   
The order of magnitude of the smallest eigenvalues of the correlation matrices for the unthinned fits is $10^{-6}$ for $w_0$,
$10^{-9}$ for $w_2$ and $10^{-12}$ for $w_3$ and $w_4$.\footnote{The smallest eigenvalue
in each case is not very sensitive to $s_0^{\rm min}$, at least in the range $s_0^{\rm min}=3.00$ to $3.55$~GeV$^2$.   The largest eigenvalue is always of order 10.}  
The smallness of these eigenvalues, 
which reflects the very strong correlations between data at different values of
$s_0$, originates in the fact that we integrate the same data to obtain all of the $I^{(w_i)}(s_0)$.
While we take the consistency of our results across the different weights (note, in particular, the
consistency for both $\a_s$ and $C_6$) as a confirmation
of the reliability of the correlated fits, it is possible that the very small eigenvalues in the
case of weights $w_3$ and $w_4$ result in somewhat larger values of $\c^2$ for these
fits, thus reducing associated $p$-values.   Indeed, we find that the fits with weights $w_3$ 
and $w_4$, for which these lowest eigenvalues are very small, improve by thinning out the
data:  $p$-values increase, while fit parameter values remain stable, for $s_0^{\rm min}=3.15$,
$3.25$ and $3.35$~GeV$^2$, as shown in Tables~\ref{tab3} and~\ref{tab4}.   Thinning by 
a factor 2 changes the lowest eigenvalues for these weights from $\sim 10^{-12}$
to $\sim 10^{-9}$.   A similar effect occurs for $s_0^{\rm min}=3.25$~GeV$^2$ and weight $w_0$,
where the lowest eigenvalue changes to $\sim 10^{-4}$.
For values of $s_0^{\rm min}$ below $3.25$~GeV$^2$, we typically find no
such clear improvement and stability, suggesting a breakdown of the theoretical representation
employed in the fits.   Indeed, already at $s_0^{\rm min}=3.15$~GeV$^2$ some instability of the fit
parameters for weights $w_3$ and $w_4$ is visible, even if $p$-values do improve.   For
the weight $w_2$, the $p$-value does not increase with thinning, for $s_0^{\rm min}=3.15$~GeV$^2$.

\begin{figure}[t!]
\vspace*{4ex}
\begin{center}
\includegraphics*[width=7cm]{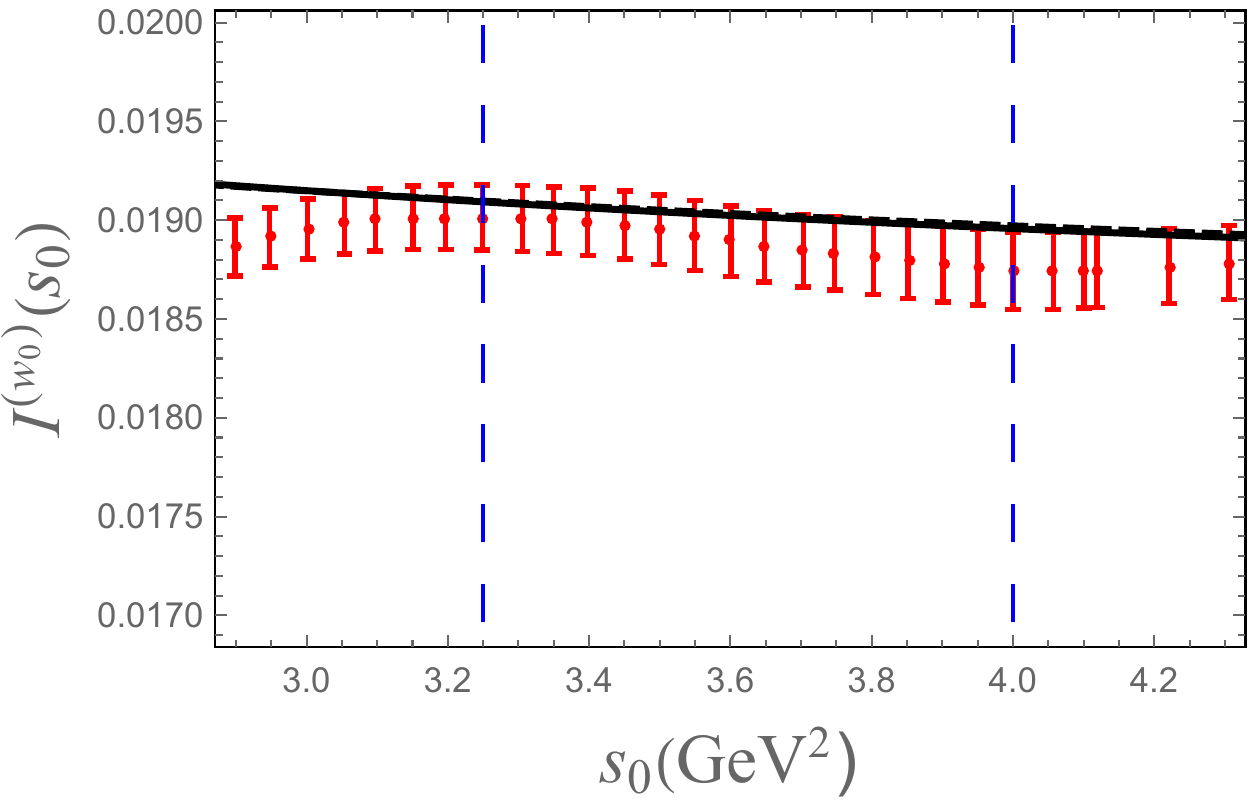}
\hspace{0.5cm}
\includegraphics*[width=7cm]{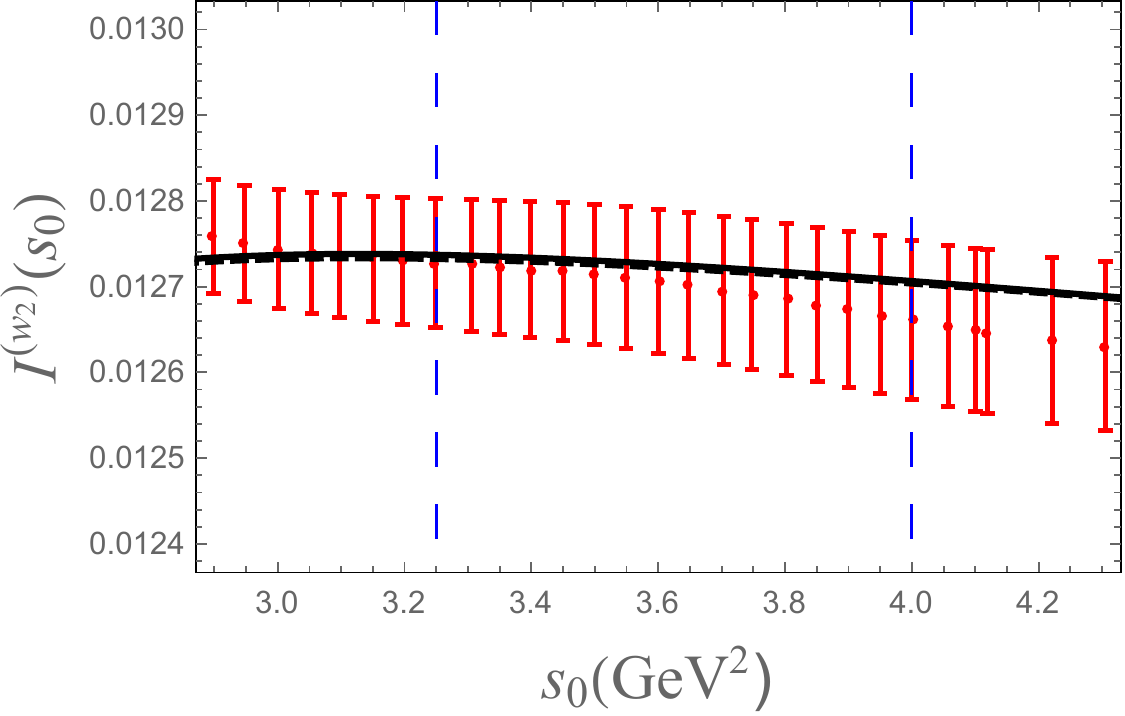}

\vspace{0.5cm}
\includegraphics*[width=7cm]{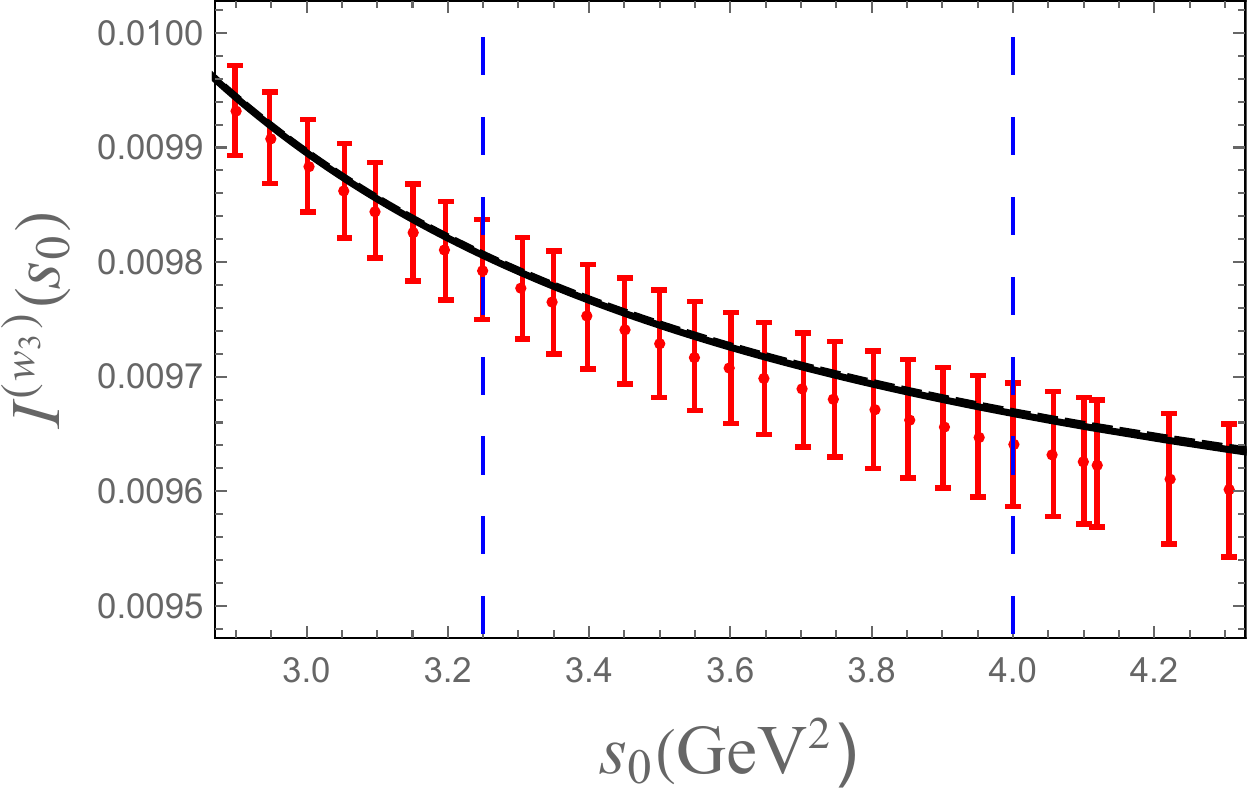}
\hspace{0.5cm}
\includegraphics*[width=7cm]{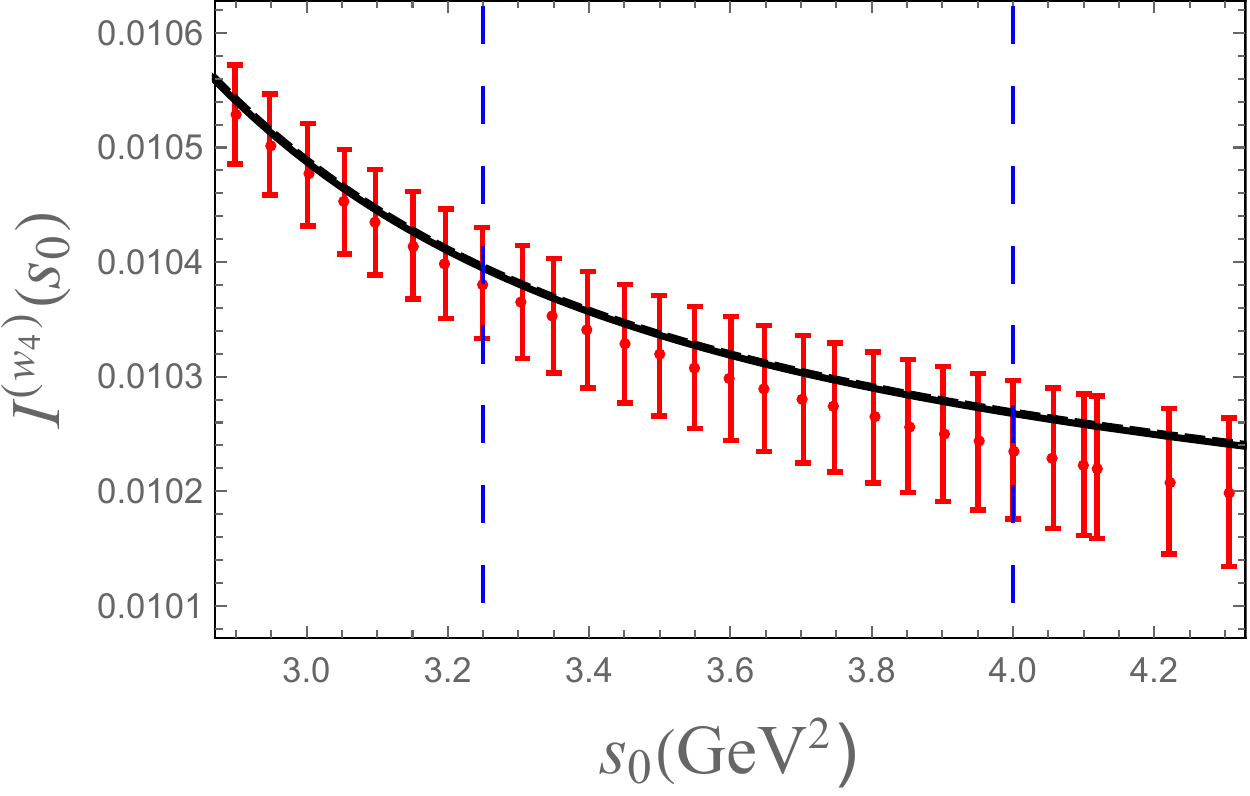}
\end{center}
\begin{quotation}
\floatcaption{fitplots}%
{{\it Comparison of the data for $I^{(w_i)}(s_0)$ with the fits on the interval $s_0^{\rm min}=3.25$ to 4~{\rm GeV}$^2$, for $i=0$ (upper left panel),
$i=2$ (upper right panel),
$i=3$ (lower left panel), and
$i=4$ (lower right panel).   Solid black curves indicate FOPT
fits, dashed curves CIPT.  The fit window is
indicated by the dashed vertical lines.  For $I^{(w_0)}(s_0)$, $I^{(w_3)}(s_0)$ and $I^{(w_4)}(s_0)$ the fit curve
is from the thinned fits in Tables~\ref{tab1}, \ref{tab3} and~\ref{tab4}, while the data for $s_0$ values 
spaced by $0.05$~{\rm GeV}$^2$are
shown.}}
\end{quotation}
\vspace*{-4ex}
\end{figure}
Based on the tables, we make the following further observations:
\begin{itemize}
\item   Fits for all weights with $s_0^{\rm min}$ values lower than those shown in the tables have
extremely small $p$-values, and these fits do not improve with thinning out the data.   
We attribute this behavior to the fact that, for such 
$s_0$, one is in the region where sizable duality violations are present in the spectrum, 
as evidenced by the peak in $R(s)$ around $s=2.8$ GeV$^2$, \seef, Figs.~\ref{Rratio} and \ref{Rblowup}.  We will return to this point in
Sec.~\ref{tests}.
\item   All FOPT fits at a given $s_0^{\rm min}$ are consistent with each other across all these tables, 
as are all CIPT fits at a given $s_0^{\rm min}$.    Note that not only the
values of $\a_s$, but also the values of $C_6$ are consistent, with $C_6$ being determined
by all fits with pinched weights.
\item   The difference between FOPT and CIPT results for $\a_s(m_\t^2)$ 
is about $0.009$ from Eq.~(\ref{alphasw0}),
about $0.007$ from Eq.~(\ref{alphasw2}), about $0.005$ from Eq.~(\ref{alphasw3}) and about
$0.006$ from Eq.~(\ref{alphasw4}).
This is much smaller than corresponding differences obtained from hadronic $\t$-decay analysis,
which are $0.022$ from the OPAL data \cite{alphas2} and $0.016$ from the ALEPH data
\cite{alphas14} (\seef\ Sec.~\ref{tau}).   The FOPT-CIPT difference is still
significant, because, for a given weight and a given $s_0^{\rm min}$, the FOPT and CIPT values of $\a_s$
are very close to $100$\% correlated.   
\item The effect of the $D=2$ term~(\ref{D=2}) is small, but not completely negligible.   Its presence
has an effect of shifting the values of $\a_s(m_\t^2)$ obtained in our fits by of order 1-2\%.
This confirms that the details of its treatment are 
indeed insignificant.
\end{itemize}

\subsection{\label{tests} Tests}
Before we use the results thus far obtained to extract a final value for $\a_s$, we perform a number of
tests probing the stability of the values reported in Eqs.~(\ref{alphasw0}) to~(\ref{alphasw4}).
The most important of these is a test for the effects of including the model for duality violations,
described in Sec.~\ref{DV}, in the fits.  

We have performed fits including Eq.~(\ref{EMansatz}), as described in Sec.~\ref{DV}.   As input 
we used the results and covariances for $\a_s$ and $I=1$ parameters $\d_1$, $\g_1$, $\a_1$ and $\b_1$ from the
$s_0^{\rm min}=1.575$~GeV$^2$, vector-channel
fit with weight $w_0$ to the ALEPH data for the non-strange
vector-channel spectral function obtained from hadronic $\t$ decays \cite{ALEPH13}, reported in
Ref.~\cite{alphas14}; the FOPT fit version of the $I=1$ spectral function predicted by this fit
is graphically shown as the orange band in Fig.~\ref{Rbreakdown}.
The fit was performed by adding a prior to our $\c^2$ function, employing the full, five-parameter covariance matrix obtained in these 
fits. The FOPT or CIPT results from Ref.~\cite{alphas14} were used, respectively, for our FOPT or CIPT fits of the $R$-ratio data.

\begin{table}[t!]
\begin{center}
\begin{tabular}{|c|c|c|c|c|c|c|}
\hline
$s_0^{\rm min}$ (GeV$^2$) & \# dofs & $\c^2$  & $p$-value  & $\a_s$ & $\d_0$ & $\a_0$ \\
\hline
\hline
2.75 & 24 & 38.6 & 0.03 & 0.285(7) & -0.41(55) & 3.90(80) \\
2.85 & 22 & 34.4 & 0.05 & 0.285(7) & -0.18(58) & 3.15(90) \\
2.95 & 20 & 25.8 & 0.17 & 0.286(7) & 0.20(57) & 2.02(94) \\
3.00 & 19 & 21.7 & 0.30 & 0.287(7) & 0.46(57) & 1.4(1.0) \\
3.15 & 16 & 17.0 & 0.39 & 0.292(8) & 1.15(60) & 1.0(1.0) \\
3.25 & 14 & 16.8 & 0.27 & 0.291(8) & 1.08(67) & 0.9(1.1) \\
3.35 & 12 & 13.2 & 0.36 & 0.292(9) & 1.23(71) & 1.1(1.0) \\
3.45 & 10 & 11.9 & 0.29 & 0.295(9) & 1.48(70) & 1.3(1.1) \\
3.55 & 8 & 11.0 & 0.20 & 0.293(9) &1.34(74) & 1.0(1.2) \\
3.60 & 7 & 8.04 & 0.33 & 0.295(9) & 1.43(72) & 1.1(1.2) \\
3.70 & 5 & 4.37 & 0.50 & 0.292(10) & 1.34(73) & 0.4(1.3) \\
3.80 & 3 & 3.97 & 0.26 & 0.292(10) & 1.31(74) & 0.4(1.4) \\
\hline
\hline
2.75 & 24 & 37.8 & 0.04 & 0.294(8) & -0.49(56) & 3.83(80) \\
2.85 & 22 & 33.8 & 0.05 & 0.295(8) & -0.30(59) & 3.12(91) \\
2.95 & 20 & 25.5 & 0.18 & 0.296(9) & 0.05(58) & 1.97(96) \\
3.00 & 19 & 21.6 & 0.25 & 0.297(9) & 0.30(58) & 1.3(1.0) \\
3.15 & 16 & 17.4 & 0.36 & 0.303(10) & 0.94(61) & 0.9(1.1) \\
3.25 & 14 & 17.1 & 0.25 & 0.302(10) & 0.85(69) & 0.8(1.1) \\
3.35 & 12 & 13.6 & 0.33 & 0.303(11) & 0.98(72) & 0.9(1.1) \\
3.45 & 10 & 12.4 & 0.26 & 0.306(11) & 1.22(73) & 1.2(1.1) \\
3.55 & 8 & 11.5 & 0.11 & 0.304(12) &1.08(76) & 0.8(1.2) \\
3.60 & 7 & 8.56 & 0.29 & 0.306(12) & 1.18(75) & 1.0(1.2) \\
3.70 & 5 & 4.84 & 0.44 & 0.302(12) & 1.09(76) & 0.2(1.3) \\
3.80 & 3 & 4.43 & 0.22 & 0.302(12) & 1.06(77) & 0.2(1.5) \\
\hline
\end{tabular}
\end{center}
\floatcaption{tab1DV}{\it Fits to $I^{(w_0)}(s_0)$
from $s_0=s_0^{\rm min}$ to $s_0=s_0^{\rm max}=4$~{\rm GeV}$^2$. FOPT results
are shown above the double line, CIPT below.    The fits include duality violations with input from the determination of Ref.~\cite{alphas14}
of the $I=1$ parameters (and $\a_s$) added as priors.}
\end{table}%

We report the results of fits including Eq.~(\ref{EMansatz}) in the $w_0$ sum rule
in Table~\ref{tab1DV}.   In this table, to save space, we do not report the
$I=1$ duality-violating parameters, but note that they are always consistent with the prior
parameter values.   We do show the values of the $I=0$ parameters $\d_0$ and $\a_0$.\footnote{Recall that in our model of Sec.~\ref{DV} we set $\g_0=\g_1$ and $\b_0=\b_1$.} 
The errors on $\a_s(m_\t^2)$ are smaller than those reported in Table~\ref{tab1}; 
the reason for this is the fact
that we added the results of Ref.~\cite{alphas14}, including the value of $\a_s$, as priors.  
Since the goal of this study is an $R(s)$-based determination of $\a_s$, the results for $\a_s$ 
reported in Table~\ref{tab1DV} are not used in fixing the central values reported in Sec.~\ref{results};
they are, instead, used only to estimate the uncertainty induced by the presence of residual duality violations on these central results.\footnote{A combined determination from these data as well as hadronic $\t$-decay data may be interesting in its own right.}

From this table, one observes that fits to much lower values of $s_0^{\rm min}$ now
have decent $p$-values, yielding values for $\a_s(m_\t^2)$ which are significantly more stable 
as a function of $s_0^{\rm min}$ than
those reported in Table~\ref{tab1}. However, the decrease of $p$-values toward
lower $s_0^{\rm min}$, 
as well as the ``wandering'' values of $\d_0$ and $\a_0$, suggest that the \ansatz~(\ref{EMansatz})
may not adequately describe duality violations for values of $s_0\,\ltap\, 3.0$~GeV$^2$.
We ascribe this to the sizable duality-violating peak around $s=2.8$~GeV$^2$ seen in 
Fig.~\ref{Rblowup}, which is a feature of the $I=0$ part of the $R$-ratio data, as it is not
seen in the $I=1$ part shown in Fig.~\ref{Rbreakdown}.   We conclude that for $I=0$, 
the asymptotic region in which Eq.~(\ref{EMansatz}) is conjectured to hold, is probably not yet
reached for $s\,\ltap\,3$~GeV$^2$.   We show the spectral function corresponding to the 
FOPT fit of Table~\ref{tab1DV} with 
$s_0^{\rm min}=3.15$~GeV$^2$ in Fig.~\ref{spectrumfit}.   This figure confirms that it is very
difficult to fit the peak around $s=2.8$~GeV$^2$ with the \ansatz~(\ref{EMansatz}),
while a reasonable representation is obtained for $s\,\gtap\,3$~GeV$^2$.\footnote{Recall
that the apparent mismatch in the inclusive region above 4~GeV$^2$ is not excluded by 
the data in that region, \seef, Sec.~\ref{inclexcl}.}  

Figure~\ref{DVs} shows the 
contributions from duality violations to weighted integrals for $w_0$ (blue dashed curve), $w_2$
(black dot-dashed curve) and $w_3$ (red solid curve), as a function of $s_0$, 
normalized to the $\a_s$-dependent
part of the integrated perturbative contribution (the difference between the full perturbation theory 
result and the parton model contribution), employing the duality-violating parameters from the FOPT, 
$s_0^{\rm min}=3.15$~{\rm GeV}$^2$ fit of Table~\ref{tab1DV}. This ratio quantifies the size of 
integrated duality violations on the scale of the $\a_s$-dependent integrated $D=0$ contributions from 
which we aim to determine $\a_s$.   This figure illustrates how pinching indeed suppresses duality 
violations,
for those values of $s$ for which the asymptotic behavior of Eq.~(\ref{EMansatz}) applies.    As we 
have seen, this appears to work reasonably well for $I=1$ (\seef, Fig.~\ref{Rbreakdown}) for
$s\,\gtap\,1.6$~GeV$^2$, but may only work for $I=0$ for $s\,\gtap\,3$~GeV$^2$.   It is clear
that the effect of pinching is significant, and more so in the region above the $\t$ mass
($s=3.157$~GeV$^2$) than below.   We note that this figure should be taken as indicative only,
because the data do not allow a full investigation of duality violations in the $I=0$ channel,
for which no information is provided by $\t$ decays.

\begin{figure}[t]
\vspace*{4ex}
\begin{center}
\includegraphics*[width=12cm]{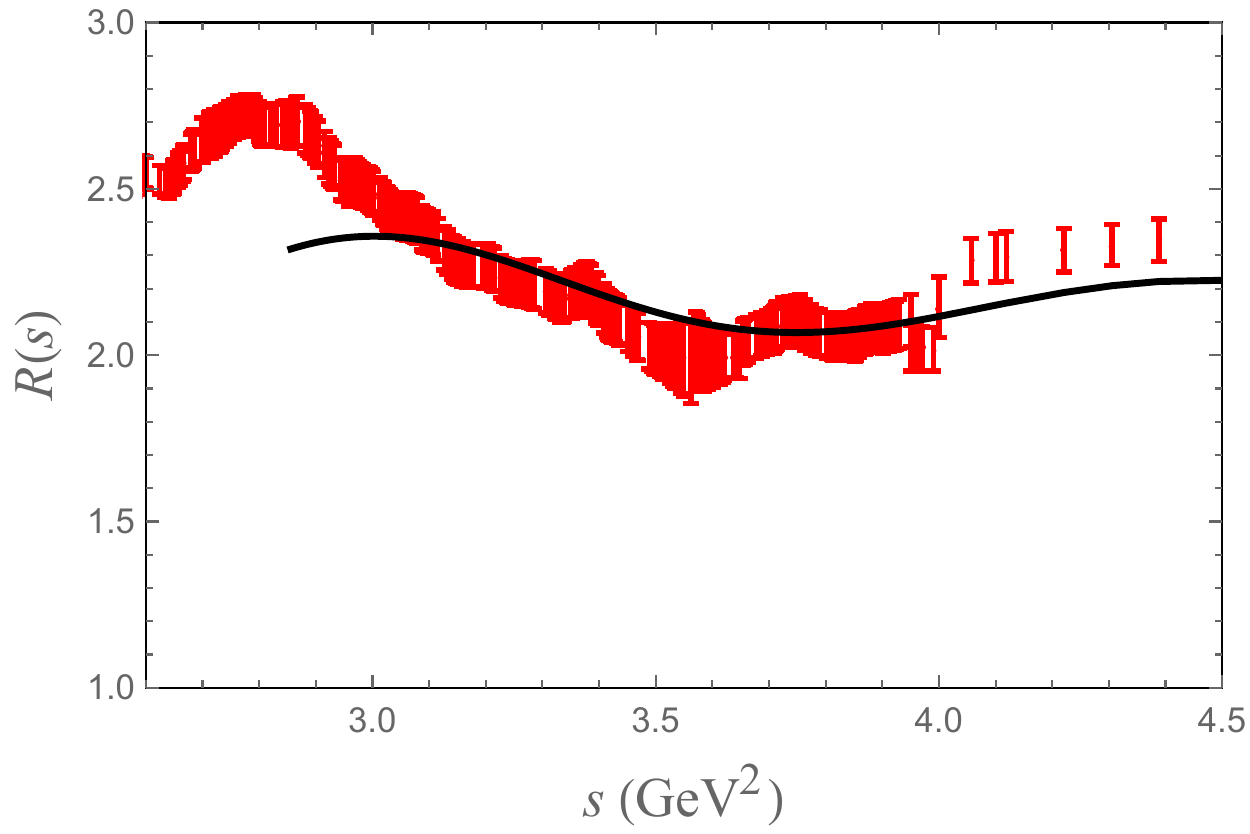}
\end{center}
\begin{quotation}
\floatcaption{spectrumfit}%
{{\it Spectral representation of the FOPT fit of $I^{(w_0)}(s_0)$ 
with $s_0^{\rm min}=3.15$~{\rm GeV}$^2$
of Table~\ref{tab1DV}.}}
\end{quotation}
\vspace*{-4ex}
\end{figure}

\begin{figure}[t]
\vspace*{4ex}
\begin{center}
\includegraphics*[width=12cm]{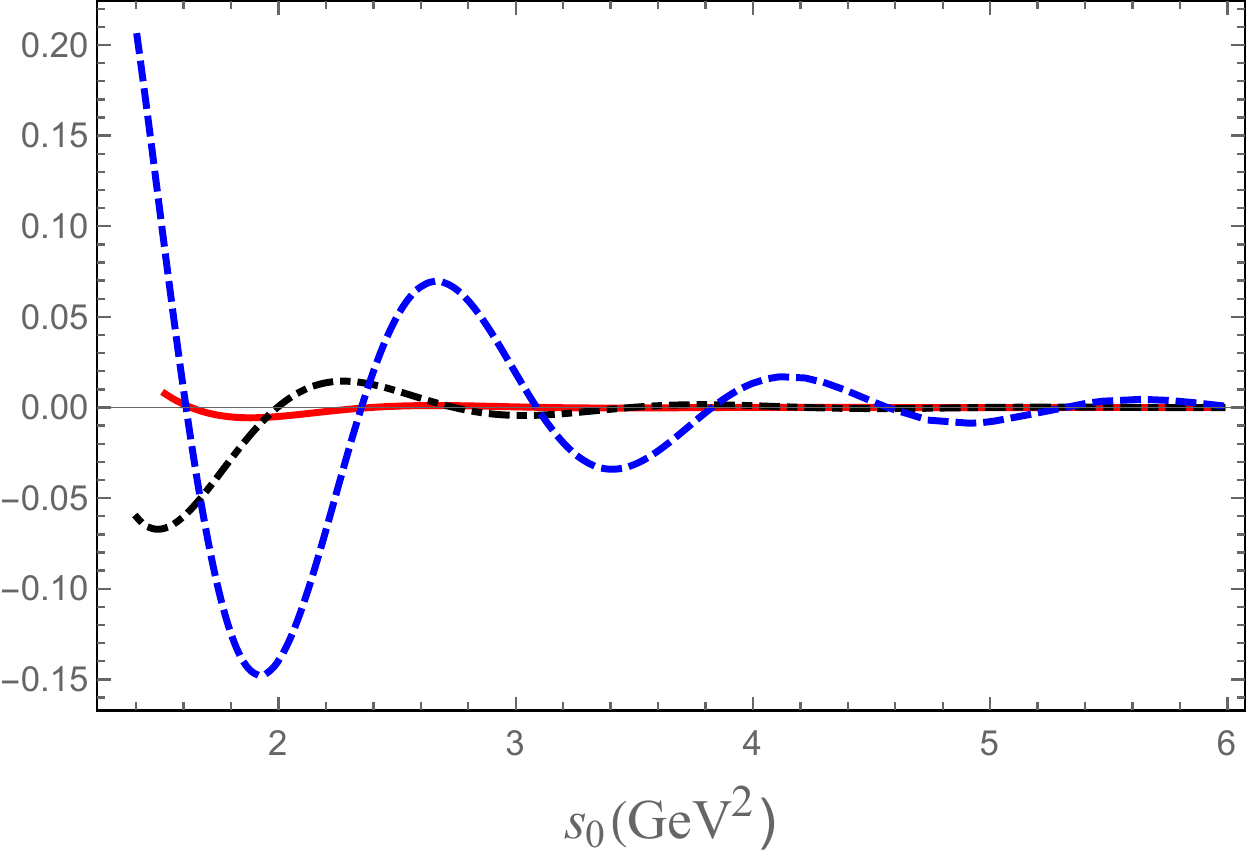}
\end{center}
\begin{quotation}
\floatcaption{DVs}%
{{\it The contribution from duality violations to the weighted spectral integrals with weights $w_0$ (blue 
dashed curve), $w_2$
(black dot-dashed curve) and $w_3$ (red solid curve), as a function of $s_0$, normalized to the
difference between perturbation theory ($D=0$ term in Eq.~(\ref{OPEdef})) and the parton model 
contribution.   The duality-violating parameters employed are those from the FOPT, $s_0^{\rm min}
=3.15$~{\rm GeV}$^2$ fit of $I^{(w_0)}(s_0)$ reported in Table~\ref{tab1DV}.}}
\end{quotation}
\vspace*{-4ex}
\end{figure}

As can be seen from the dot-dashed black and solid red curves in Fig.~\ref{DVs}, 
single-weight fits with duality violations and pinched
weights $w_{2,3,4}$ are unlikely to effectively constrain duality violations.
Nevertheless, we found that fits to $I^{(w_2)}(s_0)$ are possible, with results that
are fully compatible with Table~\ref{tab1DV} for $\a_s$, $\d_0$ and $\a_0$.  
Analogous fits for $w_3$ and $w_4$, 
for which duality violations are even more suppressed, are, unsurprisingly, not stable.

Using now the range $s_0^{\rm min}\in\{3.15,3.80\}$~GeV$^2$,
we distill the results in Table~\ref{tab1DV} into the following
estimates for $\a_s(m_\t^2)$.   We apply the same procedure as in Sec.~\ref{fits}, and find
\begin{equation}
\label{alphasw0DV}
\a_s(m_\t^2)|_{w_0}^{\rm DV}=\left\{\begin{array}{ll}
0.293(9)(2)& \qquad\mbox{(FOPT)}\ ,\cr
0.304(11)(2)& \qquad\mbox{(CIPT)}\ .
\end{array}\right.
\end{equation}

Given the caveats with our investigation of duality violations,
we use these results only to estimate the size of the systematic error associated with the presence of 
duality violations in the region above $s=3$~GeV$^2$.   
We see that (a) the value of $\a_s(m_\t^2)$ stabilizes when duality
violations are included, and (b),
that it is lower by 0.006 (0.004), for FOPT (CIPT), from comparing Eq.~(\ref{alphasw0})
with Eq.~(\ref{alphasw0DV}).
 
As an example of
the impact of integrated duality violations on FESRs involving pinched weights,
we note that, for $w_2$ and $w_3$, the maximum sizes of integrated duality violating
contributions relative to integrated $\a_s$-dependent $D=0$ terms
shown in Fig.~\ref{DVs}, in the range of $s_0$ entering the averages~(\ref{alphasw2}) 
and~(\ref{alphasw3}) are 0.3\% and 0.07\%, respectively. The maximum shift induced in
$\a_s$ at a single $s_0$ in this region is then less than 0.001 in both cases, 
much smaller than any of the other errors in the analysis.
 
We will take an error of $\pm 0.005$ as the systematic error from duality violations.    
This estimate reflects the difference between the results quoted in Eq.~(\ref{alphasw0}) and 
Eq.~(\ref{alphasw0DV}), and also safely incorporates the variations in the results reported in 
Eqs.~(\ref{alphasw0}) through~(\ref{alphasw4}).
We do not also include the second errors shown in Eqs.~(\ref{alphasw0}) through~(\ref{alphasw0DV}),
because it is very likely that the spread in values among Eqs.~(\ref{alphasw0}) through~(\ref{alphasw0DV})
is measuring essentially the same uncertainty, insofar as these second errors are due to systematic
effects.   

\begin{figure}[t!]
\vspace*{4ex}
\begin{center}
\includegraphics*[width=12cm]{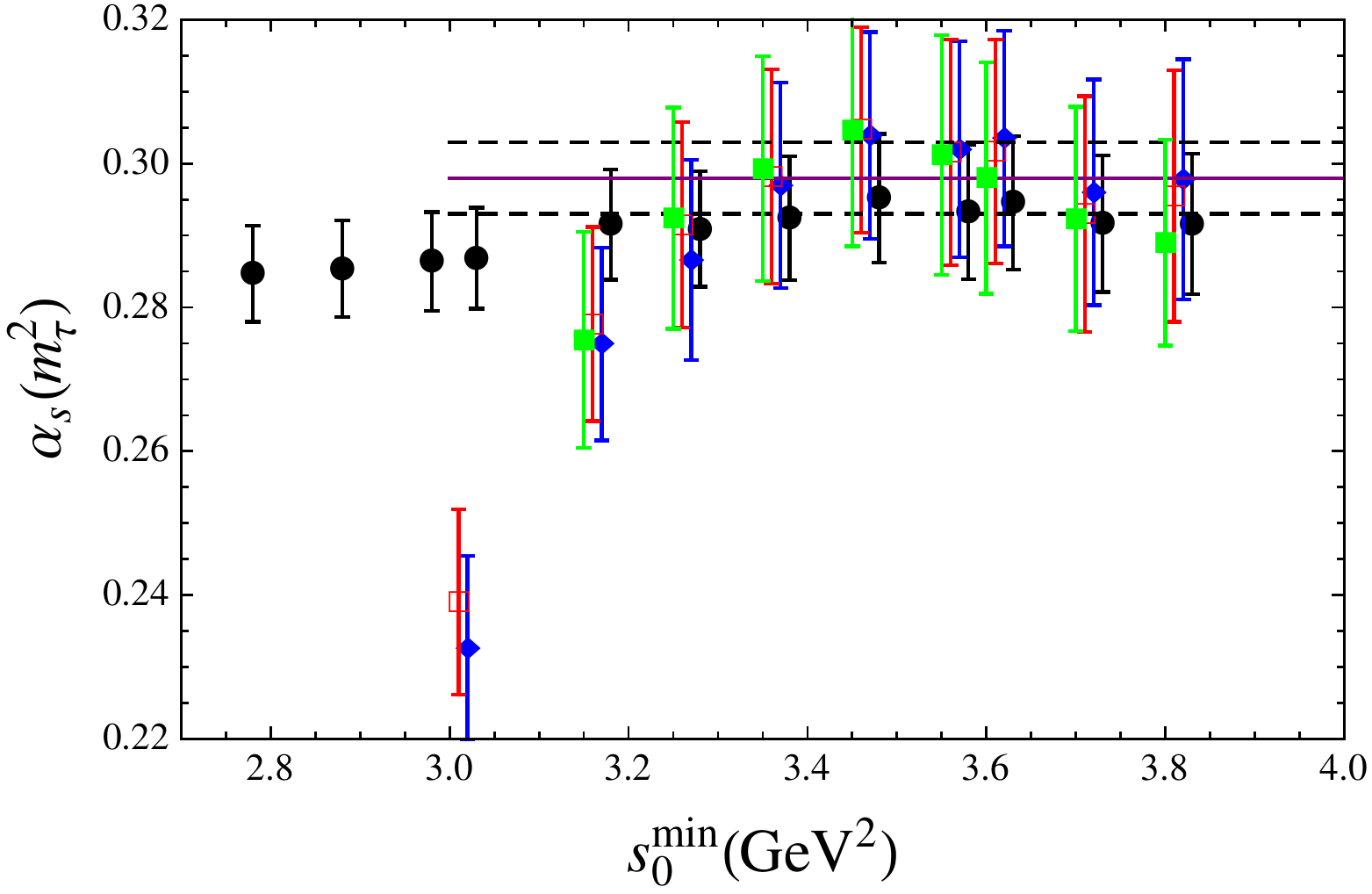}
\end{center}
\begin{quotation}
\floatcaption{alphas}%
{{\it The FOPT strong coupling $\a_s(m_\t^2)$ as a function of $s_0^{\rm min}$.   Blue data points 
(diamonds) represent values of $\a_s(m_\t^2)$ from Table~\ref{tab1}, red (open squares) those from Table~\ref{tab2},
green (filled squares) those from Table~\ref{tab3},
and black data points (filled circles) correspond to the values from Table~\ref{tab1DV}.
The solid, purple
horizontal line shows the value $0.298$, with the dashed horizontal lines showing
the values $0.298\pm 0.005$.   The red, blue and black data points have been slightly offset
horizontally for visibility.}}
\end{quotation}
\vspace*{-4ex}
\end{figure}
 
The result is illustrated in Fig.~\ref{alphas} for FOPT, 
which shows values of $\a_s(m_\t^2)$ as a function of
$s_0^{\rm min}$ from Table~\ref{tab1} (blue diamonds), Table~\ref{tab2} (red open squares), 
Table~\ref{tab3} (green filled squares), and
Table~\ref{tab1DV} (black filled circles).   Also shown is the central value for $\a_s(m_\t^2)$
obtained in Eq.~(\ref{alphasw2}) (purple horizontal line), with variations $\pm 0.005$ (dashed
horizontal lines).   The figure does not show the values reported in Table~\ref{tab4}, to avoid clutter.   
However, these additional fits do not change the picture.   For the sake of brevity we do not show the 
analogous CIPT results as these are very similar.

We investigated several other systematic issues.   One of these is the unknown value of the
perturbative six-loop Adler coefficient, $c_{51}$, for which we used an estimate $c_{51}=283$.
Varying the value of this coefficient by $\pm 283$, we find, on average, a variation of about
$\pm 0.003$ in the fitted values for $\a_s(m_\t^2)$.   We will thus allow for an additional
systematic error equal to $\pm 0.003$.

We also considered extending the range of $s_0$ values over which we fit to values larger
than 4~GeV$^2$.   We show examples of such fits of $I^{(w_0)}(s_0)$ in Table~\ref{table1csminsmax},
for both FOPT and CIPT.   The first three fits in each case have $s_0^{\rm min}$ below 4~GeV$^2$, 
in the exclusive data region, and $s_0^{\rm max}$ larger than 4~GeV$^2$, in the inclusive data region.
The other two have both $s_0^{\rm min}$ and $s_0^{\rm max}$ in the inclusive data region.  
Given the rapid decrease of integrated duality violations with
increasing $s_0$ (see Fig.~\ref{DVs}), and the fact that the impact
of integrated duality violations on $\a_s$ was already seen
to be small for the lower $s_0$ of purely exclusive region fits,
we expect such duality violating contributions to be safely
negligible for fits with both $s_0^{\rm min}$ and $s_0^{\rm max}$ in
the inclusive region, even for $w_0$.  In most cases,
indicated in the table, thinning was needed to obtain good fits.  We see that 
extending $s_0^{\rm max}$ into the inclusive region yields results in good agreement
with the results of Sec.~\ref{fits} used in the averages, with similar errors.
While the individual errors are competitive with
those in Table~\ref{tab1} when $s_0^{\rm min}<4$~GeV$^2$, the spread between different fits
becomes larger.   We should also emphasize the importance of correlations when considering
the results of these fits.   For example, taking into account correlations, we have verified that the
larger differences between the $\a_s$ values obtained with $s_0^{\rm max}=8.85$~GeV$^2$ and
$s_0^{\rm min}$ varying from $3.55$~GeV$^2$ to $4.10$~GeV$^2$ are consistent with statistical
fluctuations.

\begin{table}[t!]
\begin{center}
\begin{tabular}{|c|c|c|c|c|c|}
\hline
$s_0^{\rm min}$ (GeV$^2$) & $s_0^{\rm max}$ (GeV$^2$) & \# dofs & $\c^2$  & $p$-value  & $\alpha_s$  \\
\hline
\hline
3.25 & 4.98 & 14* & 22.5 & 0.07 & 0.297(13)   \\
3.25 & 8.85 & 26* & 32.8 & 0.17 & 0.299(13)   \\
3.55 & 8.85 & 23* & 26.8 & 0.26 & 0.310(14) \\
4.10 & 8.85 & 18* & 16.5 & 0.56 & 0.280(21) \\
6.13 & 8.85 & 15 & 15.6 & 0.41 & 0.302(24) \\
\hline
\hline
3.25 & 4.98 & 14* & 21.9 & 0.08 & 0.309(16)  \\
3.25 & 8.85 & 26* & 32.4 & 0.18 & 0.310(16)  \\
3.55 & 8.85 & 23* & 26.9 & 0.26 & 0.321(17) \\
4.10 & 8.85 & 18* & 16.1 & 0.59 & 0.288(24)  \\
6.13 & 8.85 & 15 & 14.8 & 0.46 & 0.314(28) \\
\hline
\end{tabular}
\end{center}
\floatcaption{table1csminsmax}{\it Fits to $I^{(w_0)}(s_0)$
from varying $s_0=s_0^{\rm min}$ to varying $s_0^{\rm max}$. FOPT results
are shown above the double line, CIPT below.    The fits
marked with an asterisk are thinned by a factor 2.}
\end{table}%

Similar results can be obtained for the weights $w_2$, $w_3$ and
$w_4$ and are again in good agreement with the results of Sec.~\ref{fits}, although typically for these weights thinning with a factor larger than 2 is necessary to obtain
good fits.   We therefore will only use our fits with all data in the exclusive region to obtain
our central values, considering the fits of Table~\ref{table1csminsmax} as a consistency
check.   In short, the data in the inclusive region appear to be consistent with those below
$s=4$~GeV$^2$, but with the current precision, they do not improve the accuracy in the
value of $\a_s$ that can be obtained from $R$-ratio data.   

\subsection{\label{results} Results}
Following the analysis of Secs.~\ref{fits} and~\ref{tests}, we quote as our central results for the
strong coupling from the $R$-ratio data of Ref.~\cite{KNT18} the $\overline{\rm MS}$, three-flavor values
\begin{equation}
\label{alphasfinal}
\a_s(m_\t^2)=\left\{\begin{array}{ll}
0.298\pm 0.016\pm 0.005\pm 0.003
=0.298\pm 0.017&\qquad\mbox{(FOPT)}\ ,\cr
0.304\pm 0.018\pm 0.005\pm 0.003=0.304\pm 0.019&\qquad\mbox{(CIPT)\ .}
\end{array}\right.
\end{equation}
The first error is the average fit error, the second error our estimate of the uncertainty produced by residual duality violations,
and the third error is due to the variation in $c_{51}$.   Since these errors may be considered as
independent, we combine them in quadrature to obtain our final aggregrate errors.   While we
quote values for FOPT and CIPT separately, their difference should be interpreted as another
systematic error, representing our incomplete knowledge of higher orders in perturbation theory.
While the difference, equal to 0.006, is small, it is nonetheless significant, because the FOPT and CIPT values for $\a_s(m_\t^2)$ are essentially 100\% correlated.

These three-flavor results convert to the following five-flavor results at the $Z$ mass:
\begin{equation}
\label{alphasS}
\a_s(m_Z^2)=\left\{\begin{array}{ll}
0.1158\pm 0.0022&\qquad\mbox{($\overline{\rm MS}$,\ $n_f=5$, FOPT)}\ ,\cr
0.1166\pm 0.0025&\qquad\mbox{($\overline{\rm MS}$,\ $n_f=5$, CIPT)\ .}
\end{array}\right.
\end{equation}
The central values are somewhat low compared to the PDG world average of $0.1181\pm 0.0011$ 
\cite{PDG}
and also compared to the recent high-accuracy value $0.11852\pm 0.00084$ of Ref.~\cite{alpha}, but are consistent with these results within errors.

\begin{boldmath}
\subsection{\label{tau} Comparison with the determination from hadronic $\t$ decays}
\end{boldmath}
We can also compare our results with 
those obtained from the recent analyses of OPAL and ALEPH hadronic $\t$-decay data reported in 
Refs.~\cite{alphas2,alphas14}.   A combination of these results yielded
\cite{alphas14}
\begin{equation}
\label{alphastau}
\a_s(m_\t^2)=\left\{\begin{array}{ll}
0.303\pm 0.009&\qquad\mbox{($\overline{\rm MS}$,\ $n_f=3$, FOPT)}\ ,\cr
0.319\pm 0.012&\qquad\mbox{($\overline{\rm MS}$,\ $n_f=3$, CIPT)\ .}
\end{array}\right.
\end{equation}
These values are in excellent agreement with Eq.~(\ref{alphasfinal}), differing by $0.3$, respectively,
0.7 $\s$. While the $\t$-based values have smaller total errors, we note that the difference between FOPT and CIPT values is larger
for the values obtained from $\t$ decays, in comparison with the values we obtained
here from electroproduction,
making the electroproduction-based determination more competitive with the $\t$-based determination
than the errors shown in Eqs.~(\ref{alphasfinal}) and~(\ref{alphastau}) indicate.   We also reiterate that duality violations play a significantly larger role in the $\t$-based analyses, where the sum rules are limited by kinematics to lower values of $s_0$ \cite{BGMP16}.

\vskip0.8cm
\section{\label{conclusion} Conclusion}
Recently, a new compilation of the hadronic $R$-ratio from all available experimental
data for the process $e^+e^-\to\mbox{hadrons}(\g)$ became available \cite{KNT18}.
In this paper, we used finite-energy sum rules for a determination of the strong coupling
based on these data.   

In contrast to the case of hadronic $\t$ decays,
there is no inherent limit on $s$ in $e^+e^-\to\mbox{hadrons}(\g)$, and this allowed us
to go to higher energies, where we need to rely less on models to take into account 
the non-perturbative effects associated with violations
of quark-hadron duality.   In a marked difference, only the errors in our determination,
Eq.~(\ref{alphasfinal}), required the modeling of duality violations, whereas in the case of $\t$
decays, duality violating contributions
had to be included in all self-consistent fits employed to extract $\a_s$ from the data.
Because $e^+e^-\to\mbox{hadrons}(\g)$ allowed us to probe energies above the
$\t$ mass, and because of the exponential, hence fairly rapid, decay of the strength of duality violations,
we were able to obtain stable results for $\a_s$ from sum rules which on the theory
side involve only the OPE.  This was not {\it a priori} obvious, considering that the inclusion of the
effects from duality violations has been shown to be important for the 
determination of $\a_s$ from $\t$ decays~\cite{BGMP16}.
It is thus a non-trivial result that the values for $\a_s$ we obtain from the $R$-ratio are in very
good agreement with the values for $\a_s$ obtained from $\t$ decays.   They are
also consistent within errors, when converted to values at the $Z$ mass, with the world 
average as reported in Ref.~\cite{PDG}, albeit with somewhat lower central values.   
This result provides a non-trivial test, at the current level of precision, of the perturbative running of $\a_s$ predicted by QCD even at rather low scales, a result which is far from obvious \cite{alphaPT}.

As has become common in these determinations from finite-energy sum rules,
we reported two values for $\a_s$, corresponding to two different assumptions about
how to resum unknown higher orders in perturbation theory, FOPT and CIPT.   The
difference represents our ignorance of these higher orders, assuming that, at these
energies, we have not yet reached the order in perturbation theory where its asymptotic
nature becomes manifest \cite{BJ}.   The difference between CIPT and FOPT we find from
the $R$-ratio is smaller than the one found in hadronic $\t$ decays.   It is likely that some
of this reduction can be ascribed to the extraction of $\a_s$ using sum rules at a higher
$s_0$.   However, since the
convergence properties of the perturbative expansions for the various (linear
combinations of) moments of the spectral function are not universal \cite{BBJ12},
it is not clear that a direct comparison of this difference between the
determinations from the $R$-ratio and $\t$ decays can be made.   It is for
this reason that we refrain from just adding the difference between FOPT and CIPT
as another systematic error to the total error in our determinations of $\a_s$.

Our final result, Eq.~(\ref{alphasfinal}), shows that the largest error is the fit error, which 
is experimental in nature.   This implies that more precise future data for the $R$-ratio
would help in making the determination from the $R$-ratio more precise, and provide a 
more stringent test on the workings of QCD perturbation theory at lower energies.  
The biggest impact on our determination comes from the region below 2~GeV, where
the $R$-ratio is compiled from very many carefully measured exclusive-channel 
contributions.   While much improved inclusive data in the region between 2 and 3~GeV
have more recently become available \cite{BES,BES3,KEDR}, we found that, at present, 
these inclusive data do not have much impact on the precision of our determination.  
In this respect, prospects for the release of new inclusive $R$-ratio data by 
BESIII~\cite{BESIII-newR} and the experiments at Novosibirsk (SND, CMD-3, KEDR) are 
potentially promising. 
In addition, efforts at Novosibirsk to determine the inclusive $R$-ratio at lower energies than 2 GeV~\cite{SimonRMCinclusive} would allow further study into the choices of the transition region between the sum of exclusive states and the inclusive data.

In the meantime, a project that may be worth considering 
is a determination of $\a_s$ combining hadronic $R$-ratio data and $\t$-decay data. 
Such an approach appears to be sensible in view of the consistency between our 
determinations of $\a_s$ from each of these separately.   

\vspace{3ex}
\noindent {\bf Acknowledgments}
\vspace{3ex}

We like to thank Claude Bernard and Matthias Jamin for helpful discussions.
DB, AK and KM would like to thank the IFAE at the Universitat Aut\`onoma de Barcelona
for hospitality.  The work of DB is supported by the S{\~a}o Paulo Research Foundation 
(FAPESP) Grant No. 2015/20689-9 and by CNPq Grant No. 305431/2015-3.
The work of MG is supported by the U.S. Department of Energy, Office of Science, 
Office of High Energy Physics, under Award Number DE-FG03-92ER40711.
The work of AK is supported by STFC under the consolidated grant ST/N504130/1. 
KM is supported by a grant from the Natural Sciences and Engineering Research
Council of Canada.  The work of DN is supported by JSPS KAKENHI grant numbers
JP16K05323 and JP17H01133.
SP is supported by CICYTFEDER-FPA2014-55613-P, 2014-SGR-1450. 
The work of TT is supported by STFC under the consolidated grant ST/P000290/1.


\end{document}